\documentclass[journal]{IEEEtran}
\usepackage{cite}
\usepackage{graphicx,color,overpic,psfrag}
\usepackage{subfigure}
\usepackage{amsmath}
\usepackage{times}
\usepackage{latexsym}
\usepackage{bm}
\usepackage{amssymb}
\usepackage{cases}
\usepackage{array}
\usepackage{float}
\usepackage{fancyhdr}
\usepackage{stfloats}
\usepackage{setspace}
\usepackage{multirow}
\usepackage{booktabs}
\usepackage{algorithm}
\usepackage{algorithmic}
\newcommand{\diag}{{\sf diag}}

\ifCLASSINFOpdf
\else
\fi
\begin{document}
\title{Design and Implementation of a TDD-Based 128-Antenna Massive MIMO Prototyping System}
%\title{Prototyping a TDD-Based 128-Antenna Massive MIMO System with Software Defined Radio}
%\thanks{This work is sponsored by ...}

\author{Xi~Yang,~\IEEEmembership{Student Member,~IEEE,}
        Wen-Jun~Lu,~\IEEEmembership{Member,~IEEE,}
        Ning~Wang,~\IEEEmembership{Member,~IEEE,}
        Karl~Nieman,
        Shi~Jin,~\IEEEmembership{Member,~IEEE,}
        Hongbo~Zhu,
        Xiaomin~Mu,~\IEEEmembership{Member,~IEEE,}
        Ian~Wong,
        Yongming~Huang,~\IEEEmembership{Member,~IEEE,}
        and~Xiaohu~You,~\IEEEmembership{Fellow,~IEEE}% <-this % stops a space
%\thanks{M. Shell was with the Department
%of Electrical and Computer Engineering, Georgia Institute of Technology, Atlanta,
%\thanks{Manuscript received April 19, 2016; revised August 26, 2015.}
}

\maketitle

\begin{abstract}
Spurred by the dramatic mobile IP growth and the emerging Internet of Things (IoT) and cloud-based applications,
wireless networking is witnessing a paradigm shift.
By fully exploiting the spatial degrees of freedom, the massive multiple-input-multiple-output (MIMO) technology promises significant gains
in both data rates and link reliability. %and has attracted much attention from both academia and industry.
This paper presents a time-division duplex (TDD)-based 128-antenna massive MIMO prototyping system designed to operate %at 4.1 GHz with
on a 20 MHz bandwidth.
Up to twelve single-antenna users can be served by the designed system at the same time.
System model is provided and link-level simulation corresponding to our practical TDD-based
massive MIMO prototyping system is conducted to validate our design and performance of the algorithms.
Based on the system hardware design demonstrated in this paper, both uplink real-time video and downlink data transmissions
are realized, and the experiment results show that 268.8 Mbps rate was achieved
for eight single-antenna users using QPSK modulation.
The maximum spectral efficiency of the designed system will be 80.64 bit/s/Hz by twelve single-antenna users
with 256-QAM modulation.
\end{abstract}


% \fi
\IEEEpeerreviewmaketitle

\section{Introduction}
\IEEEPARstart{O}{ver} the next five years, the global IP traffic is going to increase more than threefold
and will have achieved a hundredfold increase from 2005 to 2020 \cite{ciscoVNI}.
The increasing popularity of smart portable devices, e.g., smartphones and tablets,
and the worldwide success of the third generation (3G) and the long-term evolution (LTE) cellular standards,
are making the mobiles lead the IP traffic growth.
Behind the same rapid growth in the mobile Internet traffic, we have been seeing a dramatic change
in the growth pattern in recent years due to the rise of the machine-to-machine (M2M) type communications
and the prosper of the Internet of Things (IoT) market.
There is a demand for a redefined cellular architecture to provide good native support to
the numerous emerging applications and to improve quality-of-service (QoS) provisioning
for a large diversity of communication scenarios in future wireless networking.

The fifth generation (5G) cellular system, which is expected to be rolled out by 2020 according to the IMT-2020 road map,
is going to be a paradigm shift of mobile networking.
In order to achieve the key performance indices (KPIs) and visions of 5G,
simple evolutions from existing wireless technologies such as 3GPP LTE and Wi-Fi is not sufficient.
Disruptive new technologies on both the network side and the user side must be introduced,
among which the massive multiple-input-multiple-output (MIMO) is considered as
the most significant breakthrough in base station (BS) technologies \cite{disruptive5G}.
Different from the conventional multi-user MIMO (MU-MIMO), by using a large excess of very low power BS antennas
to serve a relatively small number of user equipments (UEs) over the same time-frequency resource block,
massive MIMO promises significant gains in wireless data rates and link reliability.
In the past few years, the massive MIMO technology has been attracting increasing attention from both academia and industry and become one of the most dynamic research topics in wireless communications  \cite{rusek2013scaling,larsson2014massive}.

It is shown in \cite{marzetta2010noncooperative} that in a time-division duplex (TDD) massive MIMO system
equipped with unlimited number of BS antennas, the MU-MIMO channel is asymptotically orthogonal
when the channel coefficients for different antenna elements are independent and identically distributed (i.i.d.).
Therefore, by fully exploiting the spatial degrees of freedom of the large-scale antenna array,
the hardware-friendly linear precoding schemes, e.g., maximal-ratio transmitting (MRT) and zero-forcing (ZF),
which can be implemented with channel state information (CSI) acquired from uplink pilot based channel estimation,
is sufficient to achieve the optimal performance asymptotically.
In addition, the work of Ngo \cite{ngo2013energy} presents that with perfect CSI, each single-antenna UE in a massive MIMO system can scale down its transmit power proportional to the number of antennas at the BS
to achieve the same uplink performance as in single-input single-output (SISO) transmissions.
Similarly, in the imperfect CSI scenario, the scaling coefficient for the UE transmit power is the square root of the number of BS antennas.
This leads to higher energy efficiency and is very important to future wireless networks where excessive energy consumption is a growing concern.
%Hence, massive MIMO has become one of the key technologies for 5G.

While the massive MIMO technology has many features desirable in future wireless networking,
the use of large-scale antenna array raises new issues that must be addressed
in the design of massive MIMO based wireless systems.
First of all, it becomes increasingly difficult to obtain accurate instantaneous CSI at transmitter (CSIT)
for the downlink, especially when the system operates in frequency-division duplex (FDD) mode.
The pilot overhead in FDD-based massive MIMO system for downlink CSIT acquisition is shown to be
proportional to the BS antenna array size \cite{rusek2013scaling}.
Moreover, even in TDD mode, there exists hardware mismatch between the BS and the UE,
which impairs channel reciprocity and necessitates calibration before downlink transmission \cite{larsson2014massive,vieira2014reciprocity,zhang2015large,vieira2016reciprocity,wei2016mutual}.
Furthermore, due to the use of large excess antenna arrays at the BSs, the hardware complexity and computational complexity are significant and they increase with the size of the antenna arrays,
which poses new challenges to the design and implementation of massive MIMO based systems in:
1) flexible software defined radio (SDR) solution to receive and send radio-frequency (RF) signals,
2) precise time and frequency synchronization among different RF devices,
3) high throughput bus to transfer and collect massive data,
and 4) high processing power required by the real-time signal processing
in the execution of physical layer (PHY) and media access control layer (MAC) functionalities.
Despite the aforementioned challenges, it is of great significance to build prototyping massive MIMO system
to help verifying its potentials and perfecting the technology before commercial deployments.
Fortunately, there are several basic prototyping works on massive MIMO such as the Argos \cite{shepard2012argos,shepard2013argosv2,shepardthesis2012argos} and the LuMaMi \cite{vieira2014flexible}.
Several leading communication network equipment manufacturers such as Huawei and Nokia Siemens Networks
have also involved in massive MIMO related research and development activities.

The Argos developed by the Rice University team in collaboration with Alcatel-Lucent
shows the feasibility of the massive MIMO concept with a 64-antenna array prototype in indoor environments.
By using hierarchical and modular design principles, Argos achieves scalability and flexibility in its implementation.
Argos V1 \cite{shepard2012argos} built a 64-antenna BS and served 15 single antenna users simultaneously through
ZF and multi-user beamforming with 0.625 MHz bandwidth in TDD mode where channel reciprocity holds.
Channel measurements were collected for both line-of-sight (LOS) and non-line-of-sight (NLOS) scenarios
and the the experimental results for cell capacity were presented.
Argos V2 \cite{shepard2013argosv2}, which is an upgraded version of Argos V1,
extended the number of BS antennas to 96 and supported 32 simultaneous data streams.
Both Argos V1 and V2 are built based on commercially available hardware, e.g.,
Wireless Open Access Research Platform (WARP) which has an open programmable FPGA and two RF chains on board.
The LuMaMi (Lund Massive MIMO) \cite{vieira2014flexible} is an SDR-based testbed employing a modular architecture
and supports up to 100 transceiver chains at the BS over 20 MHz orthogonal frequency division multiplexing (OFDM) bandwidth.
Similar to the WARP, each module in the LuMaMi system contains both RF and signal processing components.
%The LuMaMi BS is configured to support up to 100 transceiver chains,
%and orthogonal frequency division multiplexing (OFDM) with 20 MHz bandwidth is used.

In order to alleviate the overwhelming processing burden of data transmission and signal processing
due to the dramatically increased BS antenna array size, %in LuMaMi a modular hardware architecture is adopted,
distributive implementation of functionalities such as MIMO detection and precoding is adopted,
where the hardware system is divided into subsystems, i.e. smaller groups of modules,
and the processing power is evenly distributed among the subsystems.
While all data bytes to transmitted are generated at a central controller which is a separate processing unit,
the radio frequency band of an OFDM symbol is divided into several sub-bands such that each sub-band
is assigned to a subsystem which is responsible for the processing of the RF signal of the subsystem itself and the baseband processing of the sub-band of the whole system.
Specifically, for the uplink, each subsystem splits its received RF signals into sub-band signals
and distributes the sub-band signals to responsible subsystems.
After collecting the sub-band signals from all the subsystems, each subsystem conducts baseband signal processing,
whose output is then fed to central processing unit.
Similarly, in the downlink each subsystem receives baseband signal of its responsible sub-band
from the central baseband processing unit and conducts MIMO precoding.
Each subsystem then distributes the responsible processing output to other subsystems. After collecting the sub-band signals from all the subsystems, each subsystem conducts RF signal processing and performs data transmission.

%the whole system is divided into subsystems.
%Each subsystem has the same responsibility for splitting the current subsystem's whole antennas baseband data into sub-band, transferring related sub-band data to the other subsystems, and aggregating the corresponding sub-band data from the other subsystems and doing the whole system's sub-band data processing such as MIMO detecting and precoding and so on.
%Based on the system architecture and hardware implementation of LuMaMi,
%initial measured results for an over-the-air uplink massive MIMO transmission
%with spatial multiplexing of four single antenna users was performed
%with all basedband processing conducted at the central controller.}}

However, existing works on prototyping massive MIMO have limitations in terms of
suitability for commercial implementations.
Firstly, although both Argos and LuMaMi are operated in TDD mode, there is urgently lack of a clear link-level procedure illustration for TDD data transmission corresponding to the practical testbeds.
Secondly, only 0.625 MHz bandwidth downlink channel with 64 BS antennas and 100-antenna MIMO uplink over 20 MHz bandwidth are realized by Argos and LuMaMi, respectively, which is insufficient for real world massive MIMO deployments.
In addition, the communication air interface configuration with TDD, including initial RF chain calibration,
uplink channel estimation from uplink pilot symbol, uplink data transmission, TDD switching, downlink precoding,
downlink pilot and data transmission etc. are not available.
Hence, there is a demand for continued effort on massive MIMO prototyping testbed development from both academia and industry.
%which can be used to further validate the massive MIMO concept with over-the-air real-time data transmission and even larger numbers of antennas.

In this paper, we try to resolve the aforementioned limitations of existing work and present design and implementation
of a TDD-based 128-antenna massive MIMO prototyping system based on SDR platform.
%The system operates at 4.1 GHz with 20 MHz bandwidth.
The designed system can serve up to twelve single antenna users on the same frequency-time resource block
and a resource demanding video streaming service is used to test our design.
Similar to Argos and LuMaMi, because of flexibility and scalability considerations, we have divided our 128 antenna massive MIMO prototyping system into subsystems with each subsystem consisting of 16 antennas (8 modules).
In order to improve the computational capability to handle the massive baseband data (about 6.45 GB/s),
four FPGA co-processors are used in the system.
Moreover, both the hardware and the software utilized by our system are built with
commercially available products/solutions, which makes our system stable, friendly for customization,
and sufficient accurate.
The main contributions of this work are summarized in the following:
\begin{enumerate}
  \item[a)] A link-level communication procedure illustration and system simulation of our TDD-based 128-antenna massive MIMO prototype design are presented.
      Hardware mismatch between the BS and UEs for the uplink and downlink channel measurements
      is considered in the link-level simulation.
%      Due to the elaborative consideration of the practical hardware effects and the consistency with the real link-level processing procedure,
      The emulation of the real link-level processing procedure for the prototyping system is therefore
      of great value to evaluation of the processing algorithms implemented,
      e.g., reciprocity calibration and multi-user precoding.
  \item[b)] We have designed and built a practical TDD-based 128 antenna massive MIMO prototyping system,
      which realizes eight single-antenna users' real-time over-the-air uplink and downlink data transmission
      based on air interface synchronization through primary synchronization signal (PSS).
      %the maximal spectral efficiency can be achieved by the usage of 256-QAM and twelve single-antenna users is 80.64bit/s/Hz .
  \item[c)] The real-time measurements of the 20 MHz bandwidth multi-user massive MIMO channel are presented,
      and the impact of the reciprocity calibration is also studied.
\end{enumerate}

%\IEEEPARstart{T}{his} demo file is intended to serve as a ``starter file''
%for IEEE journal papers produced under \LaTeX\ using
%IEEEtran.cls version 1.8b and later.
%% You must have at least 2 lines in the paragraph with the drop letter
%% (should never be an issue)
%I wish you the best of success.

%(1) The concept of massive MIMO from Mazetta's paper.
%(2) Brief introduction of Argos, Samsung, and LuMaMi.
%(3) Point out the main features of our 128 antenna massive MIMO system.

The rest of this paper is organized as follows.
To give a clear overview of the practical massive MIMO system in principle,
the theoretical system model is given in Section II.
Section III describes the system link-level simulation in detail.
The system design and experiment setup for validating our prototype design are presented in Section IV,
and Section V presents the corresponding experimental results.
Concluding remarks are given in Section VI.

\emph{Notation:} We use uppercase and lowercase boldface letters to denote matrices and vectors, respectively.
The $N \times N$ identity matrix is denoted by ${\bf{I}}_N $, the all-zero matrix is denoted by ${\bf{0}}$,
the all-one matrix is denoted by ${\bf{1}}$.
${{\mathbf{e}}_k} \in {\mathbb{R}^{K \times 1}}$ represents the $k$th unit vector, i.e.,
the vector which is zero in all entries except the $k$th entry which it is set to $1$.
The superscripts $(\cdot)^{H}$, $(\cdot)^{T}$, and $(\cdot)^{*}$ stand for the conjugate-transpose, transpose,
and conjugate operations, respectively.

%%%%%%%%%%%%%%%%%%%%%%%%%%%%%%%%%%%%%%%%%%%%%%%%%%%%%%%%%%%%%%%%%%%%%%%%%%%%%%%%%%%%%%%%%%%%%
\section{System Model}
%In this section, the theoretical system model corresponding to our practical TDD-based massive MIMO prototyping system is provided for the purpose of offering a better understanding of the prototyping system in principle.
In this section, the analytical model for TDD-based massive MIMO system is presented.
This offers an overview of the fundamentals and basis behind the proposed prototyping system design,
which helps the readers to better understand our design principles and the subsequent results.

\subsection{Communication Scenario}
\begin{figure}[!t]
\centering
\includegraphics[width=2.5in]{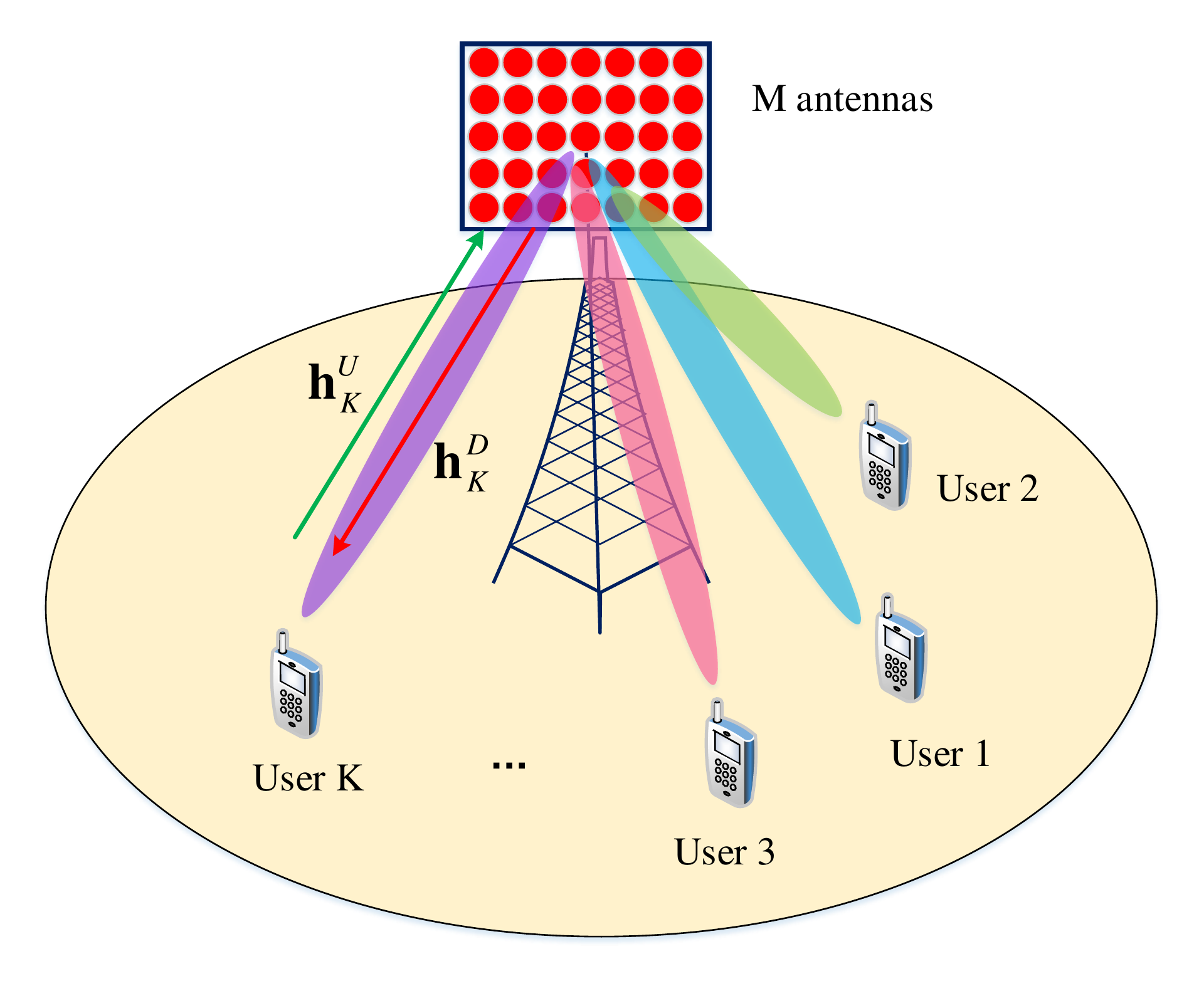}
\caption{A single-cell multi-user (MU) massive MIMO system. BS is equipped with $M$ antennas and simultaneously serves $K$ ($K \ll M$) randomly distributed single-antenna users in the same time-frequecny resource.}
\label{fig:1}
\end{figure}

\begin{figure}[!t]
\centering
\includegraphics[width=0.5\textwidth]{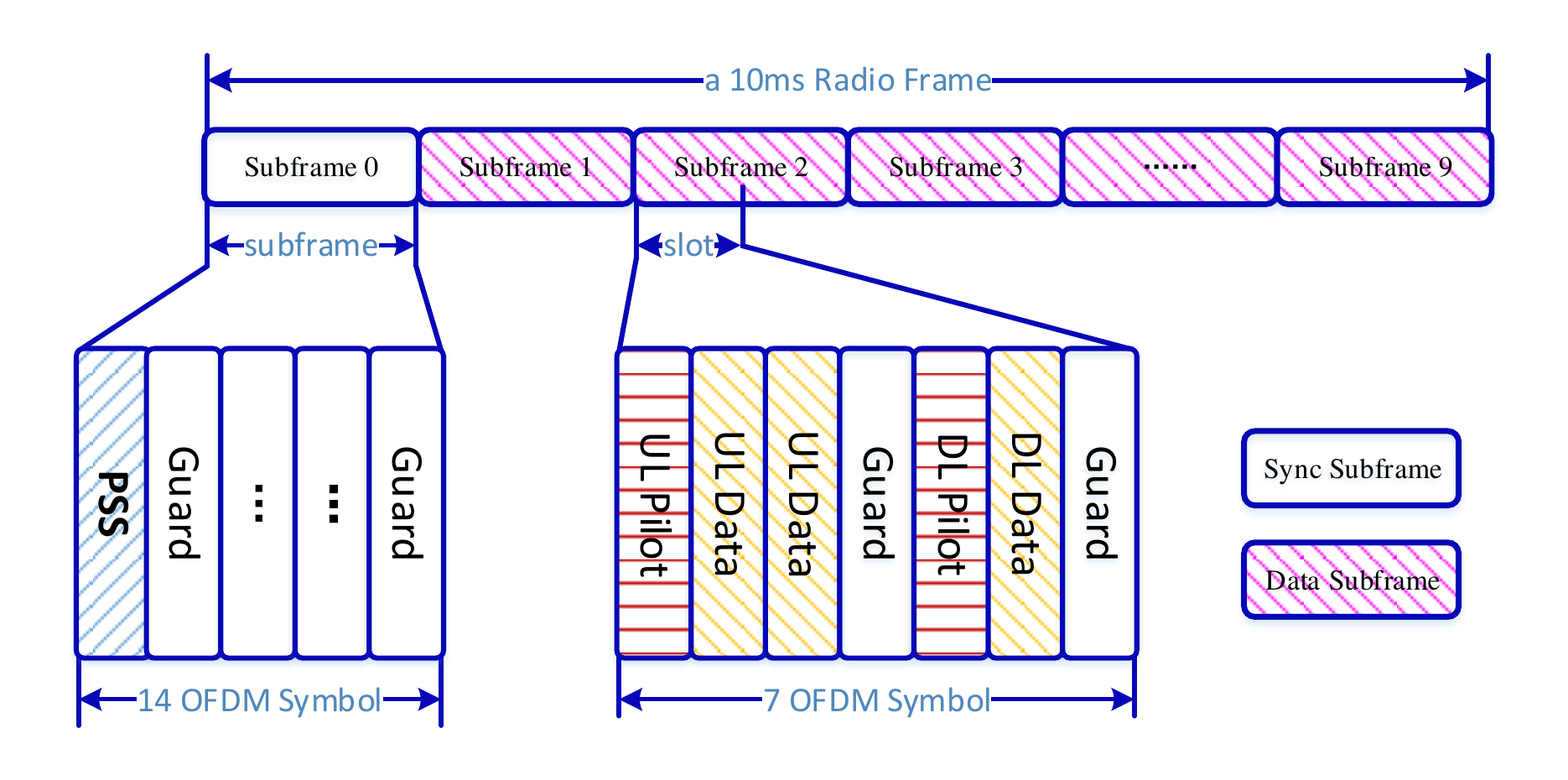}
\caption{Frame structure. A 10 ms radio frame has been divided into 10 subframes. Subframe 0 is used for synchronization between the BS and UEs, Subframe 1 to Subframe 9 are used for data transmission, where each subframe is composed of two 0.5 ms slots and each slot consists of seven OFDM symbols: Uplink (UL) Pilot, UL Data, UL Data, Guard, Downlink (DL) Pilot, DL Data, and Guard.}
\label{fig:2}
\end{figure}

Consider a single-cell multi-user (MU) massive MIMO system adopting $N$-subcarrier OFDM.
The base station is equipped with $M$ antennas and simultaneously serves $K$ ($K \ll M$) randomly distributed single-antenna users in the same time-frequency resource block, as shown in Fig. \ref{fig:1}.
The system operates in TDD mode, and Fig. \ref{fig:2} is the frame structure adopted.
%In the frame structure,
Specifically, a 10 ms radio frame is divided into 10 subframes.
Except for Subframe 0, which is used for synchronization between the BS and UEs through PSS,
all the other subframes, i.e. Subframe 1 through Subframe 9, are used for data transmission
and have the same structure of two 0.5 ms time slots.
Each time slot consists of seven OFDM symbols as shown in Fig. \ref{fig:2} and the configuration
can be adjusted flexibly to meet the data transmission demand in different scenarios.
%i.e. Uplink (UL) Pilot, UL Data, UL Data, Guard, Downlink (DL) Pilot, DL Data, and Guard.
Note that the Guard symbol is reserved for TDD switch.
%and the frame structure can be adjusted flexibly to meet the data transmission demand in different scenarios.
In this example, we allocate two OFDM data symbols for the uplink because of
the real-time video streaming application to be used to test the prototyping system.

\subsection{Uplink Data Transmission}
Before the uplink data transmission, uplink pilot symbol is firstly transmitted by the $K$ single-antenna users (labeled as ${\sf UE}_0, \dots, {\sf UE}_{K-1}$) for channel measurements at the BS.
We adopt frequency orthogonal pilots for different users and zero-hold method
is utilized in the least square (LS) channel estimation.
The $K$ neighbouring subcarriers are successively allocated to $K$ users, e.g., Subcarrier $0$ for ${\sf UE}_0$, Subcarrier $1$ for ${\sf UE}_1$, \dots,  Subcarrier $K-1$ for ${\sf UE}_{K-1}$, Subcarrier $K$ again for ${\sf UE}_0$ and so forth.
Details about the resource allocation scheme is illustrated in Section III.
Furthermore, we group every $K$ consecutive subcarriers into a sub-band
which gives $N/K$ sub-bands marked as Sub$0$, Sub$1$, \dots and Sub$N/K-1$.
Each user only transmits pilot on its pilot subcarrier and the other $K-1$ subcarriers on the sub-band are preserved.
Once the channel is estimated for ${\sf UE}_k$, $k=0\dots K-1$ by the pilot on Subcarrier $n$ in Sub$i$,
i.e. $n \in [iK,iK + K - 1]$, the channel estimate will be regarded as the same for ${\sf UE}_k$ in Sub$i$.
The signal model of the uplink pilot transmission in Sub$i$ over block fading channel is given by
\begin{equation}\label{eq:uplink pilot model}
{{\bf{R}}_i} = {{\bf{\bar H}}_i^U}{{\bf{P}}_i} + {\bf{Z}},
\end{equation}
where ${{\bf{R}}_i} \buildrel \Delta \over = [{{\bf{r}}_{iK}},{{\bf{r}}_{iK + 1}}, \ldots ,{{\bf{r}}_{iK + K - 1}}]\in{\mathbb{C}^{M \times K}}$ is the received signal at the BS,
${{\bf{\bar H}}_i^U} \buildrel \Delta \over = [{{\bf{\bar h}}_{iK}^U},{{\bf{\bar h}}_{iK + 1}^U}, \ldots ,{{\bf{\bar h}}_{iK + K - 1}^U}]\in{\mathbb{C}^{M \times K}}$ denotes the uplink channel in Sub$i$,
${{\bf{P}}_i} \buildrel \Delta \over = \diag({p_{iK,0}},{p_{iK + 1,1}}, \ldots {p_{iK + K - 1,K - 1}})\in{\mathbb{C}^{K \times K}}$ is the pilot matrix
with $j_{th}$ column ($j=0,1,\ldots,K-1$) representing the $j_{th}$ user's uniform pilot vector,
${\bf{Z}} \in{\mathbb{C}^{M \times K}}$ is the noise matrix where the elements of ${\bf{Z}}$
are i.i.d.~complex Gaussian noises with $E[{\bf{Z}}{{\bf{Z}}^H}] = {\sigma ^2}{\bf{I}}$.
Therefore, by using the LS channel estimation and zero-hold method,
the multi-user uplink channel on Subcarrier $n$ in Sub$i$ can be given as
\begin{equation}\label{eq:uplink channel estimate1}
{{{\bf{\hat H}}}_n^U} = {{\bf{R}}_i}{\bf{P}}_i^*,\quad \forall n \in [iK,iK + K - 1],
\end{equation}
Substitute (\ref{eq:uplink pilot model}) into (\ref{eq:uplink channel estimate1}), we obtain
\begin{equation}\label{eq:uplink channel estimate2}
{\bf{\hat H}}_n^U = {\bf{\bar H}}_i^U + {\bf{\tilde Z}},\quad \forall n \in [iK,iK + K - 1],
\end{equation}
where ${\bf{\tilde Z}} = {\bf{ZP}}_i^*$.
Note that the MU-MIMO uplink channel estimation is the same for all the subcarriers in the same sub-band.

According to the frame structure shown in Fig. \ref{fig:2}, after the uplink pilot symbol,
uplink data symbols are transmitted by the $K$ users over the same time-frequency resource blocks.
Assume ${\bf s}_n \triangleq \left[{s_{n1}}, \dots ,{s_{nK}} \right]^T \in \mathbb{C}^{K \times 1}$
is the transmitted vector with $s_{nk}$, ($k=1,\ldots,K$), being the zero-mean transmitted
complex confidential message to ${\sf UE}_k$, and all $s_{nk}$s are i.i.d. with unit variance,
the received signal at the BS is given by
\begin{equation}\label{eq:uplink received signal}
{{\bf{y}}_n^U} = {\bf{H}}_n^U{{\bf{s}}_n} + {{\bf{z}}_n},
\end{equation}
where ${\bf{H}}_n^U$ is a $M\times K$ channel matrix on subcarrier $n$,
and ${{\bf{z}}_n}$ is a complex Gaussian noise vector with i.i.d. elements
and $E[{{\bf z}_n}{{{\bf z}_n}^H}] = {\sigma ^2}{\bf{I}}$.
Using the pilot-based channels estimation, the linear minimum mean square error (LMMSE) detector
is implemented at the base station.
The detected signal vector is therefore given by
\begin{equation}\label{eq:uplink estimated signal}
{{{\bf{\hat s}}}_n} = {{\bf{W}}_{lmmse,n}}{{\bf{y}}_n^U},
\end{equation}
where
\begin{equation}\label{eq:W lmmse}
{{\bf{W}}_{lmmse,n}} = {[{({\bf{\hat H}}_n^U)^H}{\bf{\hat H}}_n^U + {\sigma ^2}{{\bf{I}}_K}]^{ - 1}}{({\bf{\hat H}}_n^U)^H}.
\end{equation}

As can be observed from (\ref{eq:W lmmse}), matrix inversion must be implemented to achieve the above LMMSE detector,
which results in significant computational and hardware complexity.
This may be impractical in practical system, especially when large-scale channel matrices come into play in massive MIMO.
As a consequence, we use the QR decomposition approach as discussed in \cite{FEdman,wubben2003mmse,MMyllyla,MKarkooti}
to solve the matrix inversion problem.
We assume the extended channel matrix is defined as
\begin{equation}\label{eq:B}
{\mathbf{B}} = \left[ {\begin{array}{*{20}{c}}
  {{\mathbf{\hat H}}_n^U} \\
  {\sigma {{\mathbf{I}}_K}}
\end{array}} \right] = {\mathbf{Q\underset{\raise0.3em\hbox{$\smash{\scriptscriptstyle-}$}}{R} }} = \left[ {\begin{array}{*{20}{c}}
  {{{\mathbf{Q}}_1}} \\
  {{{\mathbf{Q}}_2}}
\end{array}} \right]{\mathbf{\underset{\raise0.3em\hbox{$\smash{\scriptscriptstyle-}$}}{R} }} = \left[ {\begin{array}{*{20}{c}}
  {{{\mathbf{Q}}_1}{\mathbf{\underset{\raise0.3em\hbox{$\smash{\scriptscriptstyle-}$}}{R} }}} \\
  {{{\mathbf{Q}}_2}{\mathbf{\underset{\raise0.3em\hbox{$\smash{\scriptscriptstyle-}$}}{R} }}}
\end{array}} \right],
\end{equation}
where QR decomposition is introduced in the second equation and the $(M+K) \times K$ matrix ${\bf{Q}}$ with orthonormal columns was partitioned into the $M \times K$ matrix ${\bf{Q}}_1$ and $K \times K$ matrix ${\bf{Q}}_2$, ${{\mathbf{\underset{\raise0.3em\hbox{$\smash{\scriptscriptstyle-}$}}{R} }}}$ is a $K \times K$ upper triangle matrix. Substituting (\ref{eq:B}) into (\ref{eq:W lmmse}), ${{\bf{W}}_{lmmse,n}}$ can be derived as
\begin{equation}\label{eq:W lmmse B}
{{\bf{W}}_{lmmse,n}} = {{\mathbf{\underset{\raise0.3em\hbox{$\smash{\scriptscriptstyle-}$}}{R} }}}^{-1}({{{\mathbf{\underset{\raise0.3em\hbox{$\smash{\scriptscriptstyle-}$}}{R} }}}^H})^{-1}{({\bf{\hat H}}_n^U)^H}.
\end{equation}
From (\ref{eq:B}), we can see
\begin{equation}\label{eq:Q1}
{({\bf{\hat H}}_n^U)^H} = {\bf{Q}}_1{{\mathbf{\underset{\raise0.3em\hbox{$\smash{\scriptscriptstyle-}$}}{R} }}},
\end{equation}
and
\begin{equation}\label{eq:Q2}
{\sigma {{\mathbf{I}}_K}} = {\bf{Q}}_2{{\mathbf{\underset{\raise0.3em\hbox{$\smash{\scriptscriptstyle-}$}}{R} }}},
\end{equation}
which means
\begin{equation}\label{eq:R inverse}
{{\mathbf{\underset{\raise0.3em\hbox{$\smash{\scriptscriptstyle-}$}}{R} }}}^{-1} = \frac{1}{\sigma }{\bf{Q}}_2,
\end{equation}
Combining (\ref{eq:W lmmse B}) (\ref{eq:Q1}) and (\ref{eq:R inverse}), we have
\begin{equation}\label{eq:W lmmse2}
{{\bf{W}}_{lmmse,n}} = \frac{{\bf{Q}}_2{\bf{Q}}_1^H}{\sigma }.
\end{equation}
Therefore, the matrix inversion can be replaced with the QR decomposition of the extended channel matrix $\mathbf{B}$
which can be easily realized by the means of Schmidt orthogonalization.

\subsection{Downlink Data Transmission}
In the downlink, pilot symbol is transmitted at the beginning of the time slot, followed by downlink data symbols.
The allocation of the downlink pilot is the same as in the uplink, which is to allocate orthogonal pilots in the frequency domain for different users.
Note that both the downlink pilot symbol and the downlink data symbols are precoded
by the precoding matrix in the downlink transmission.
Let ${\bf x}_n \in \mathbb{C}^{K \times 1}$ be the transmitted vector of information-bearing signals
for the $K$ single-antenna terminals satisfying $E\{ {{{\left\| {{{\bf{x}}_n}} \right\|}^2}} \} = {\rho}$.
${{\bf F}_n}\in \mathbb{C}^{M \times K}$ represents the precoder matrix.
The received signals at the $K$ users is then given by
\begin{equation}\label{eq:y downlink}
{\bf{y}}_n^D = {{\bf{H}}_n^D}{{\bf{D}}_n}{{\bf{F}}_n}{{\bf{x}}_n} + {\bf{n}},
\end{equation}
where ${{\bf{H}}_n^D}\in \mathbb{C}^{K \times M}$ is the downlink channel matrix on subcarrier $n$,
${{\bf{D}}_n} $ is a diagonal matrix with its main diagonal elements being reciprocity calibration coefficients,
and ${\bf{n}}$ is a complex Gaussian noise vector with i.i.d. unit variance elements.
The details of the reciprocity calibration coefficients are illustrated in Section IV.

In our massive MIMO prototyping system, there are two precoding algorithms employed,
the LMMSE precoding and the maximal-ratio transmitting (MRT) precoding.
For LMMSE precoding, the precoder matrix ${{\bf F}_n}$ is given by
\begin{equation}\label{eq:F precoding matrix1}
{{\bf F}_n}={{\bf{W}}_{lmmse,n}^H}{{\bf{\Lambda }}_{1n}},
\end{equation}
and for MRT, the precoder matrix ${{\bf F}_n}$ is defined as
\begin{equation}\label{eq:F precoding matrix2}
{{\bf F}_n}={({\bf{\hat H}}_n^U)^H}{{\bf{\Lambda}}_{2n}}.
\end{equation}
where the diagonal matrix ${{\bf{\Lambda}}_{1n}}$ and ${{\bf{\Lambda}}_{2n}}$ are introduced to
normalize the columns of  ${{\bf{W}}_{lmmse,n}^H}$ and ${({\bf{\hat H}}_n^U)^H}$, respectively,
and they must comply with the power constraints.

On the user side, least square channel estimation and maximal-ratio combining are employed.
Due to the use of frequency orthogonal pilots and supposing $n\mod K=k$,
the downlink pilot vector on subcarrier $n$ before precoding at the base station is given by
${{\mathbf{p}}_n} = {p_{nk}}{{\mathbf{e}}_k} \in {\mathbb{C}^{K \times 1}}$,
where ${p_{nk}}$ is a QPSK modulated symbol with unit norm for user $k$ on subcarrier $n$.
We define ${{\mathbf{D}}_n}{{\mathbf{F}}_n} \triangleq {{\mathbf{A}}_n} = [{{\mathbf{a}}_{n1}},{{\mathbf{a}}_{n2}}, \ldots ,{{\mathbf{a}}_{nK}}]$ and ${\mathbf{H}}_n^D = {[{({\mathbf{h}}_{n1}^D)^T},{({\mathbf{h}}_{n2}^D)^T}, \ldots ,{({\mathbf{h}}_{nK}^D)^T}]^T}$, where ${{\mathbf{a}}_{nk}} \in {\mathbb{C}^{M \times 1}}$ is the column vector of the effective precoder matrix ${{\mathbf{A}}_n}$,
and ${\mathbf{h}}_{nk}^D \in {\mathbb{C}^{1 \times M}}$ is the downlink channel of user $k$ on subcarrier $n$.
The estimate of the effective channel on subcarrier $n$ for user $k$ is given by
\begin{equation}\label{eq:effective channel estimate}
{{\tilde h}}_{nk}^D  = {\bf{h}}_{nk}^D{{\bf{a}}_{nk}},
\end{equation}

Subsequently, the single-antenna user processes its received signal by
multiplying the conjugate-transpose of the effective channel estimate, which, according to (\ref{eq:y downlink}), gives
\begin{equation}\label{eq:x estimate1}
{{{{\hat x}}}_{nk}} = {(\tilde h_{nk}^D)^H}{\bf{h}}_{nk}^D\sum\limits_{i = 1}^K {{x_{ni}}{{\bf{a}}_{ni}}}.
\end{equation}
Combining (\ref{eq:effective channel estimate}) and (\ref{eq:x estimate1}), we obtain
\begin{equation}\label{eq:x estimate2}
{{{{\hat x}}}_{nk}} = {\left\| {{\bf{h}}_{nk}^D{{\bf{a}}_{nk}}} \right\|^2}{x_{nk}} + {({\bf{h}}_{nk}^D{{\bf{a}}_{nk}})^H}{\bf{h}}_{nk}^D\sum\limits_{i = 1,i \ne k}^K {{x_{ni}}{{\bf{a}}_{ni}}},
\end{equation}
where the first term corresponds to the desired signal and the second term is the interference from the other users.

%%%%%%%%%%%%%%%%%%%%%%%%%%%%%%%%%%%%%%%%%%%%%%%%%%%%%%%%%%%%%%%%%%%%%%%%%%%%%%%%%%%%%%%%%%%%%
\section{Link-Level Simulation}
Link-level simulation of the TDD-based 128 antenna massive MIMO system is presented in this section.
Firstly, we give the simulation parameters including system configuration, channel parameters and resource grids of UEs.
Then we show the system block diagram which illustrates the link-level transmission procedure in detail.
Numerical results are presented at last.

\subsection{Simulation Parameters}
\begin{figure}[!t]
\centering
\includegraphics[width=0.5\textwidth]{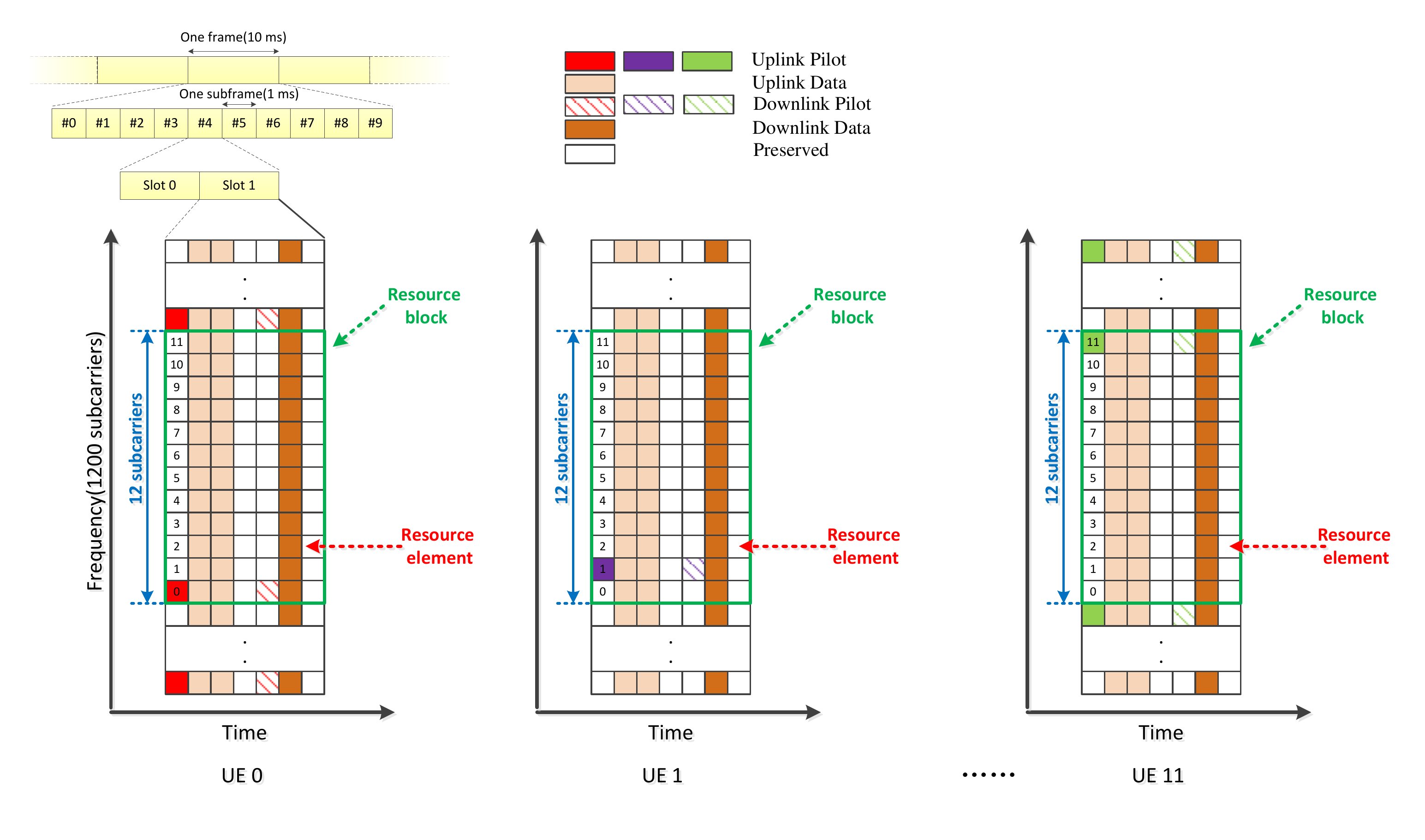}
\caption{The time-frequency resource grids of twelve single-antenna UEs. Frequency orthogonal pilots are employed for the twelve single-antenna UEs and the number of subcarriers in one subband is $12$.}
\label{fig:3}
\end{figure}

The simulation is conducted by using system and environment settings similar to the LTE cellular systems,
which sets simulation parameters as shown in Table \ref{tab:system simulation parameters}.
Both the OFDM and frequency orthogonal pilots are employed in the link-level simulation,
the time-frequency resource grids of the 12 single-antenna users are illustrated in Fig. \ref{fig:3}
which are consistent with the descriptions in Section II with each sub-band containing 12 subcarriers.
In addition, the simulation is conducted for spatial channel model (SCM) \cite{huang2004spatial}
and the settings of the channel model are given in Table \ref{tab:channel model parameters}.
\begin{table}[h]
\caption{system simulation parameters}\label{tab:system simulation parameters}
  \centering
  \begin{tabular}{lll}
    \toprule[1.2pt]
    \bf{Parameter} & \bf{Variable} & \bf{Value} \\
    \hline
    \# of BS antennas & $M$ & 128 \\
    \# of single-antenna UEs & $K$ & 12 \\
    Bandwidth  & $W$ & 20MHz \\
    Sampling Rate  & $F_s$ & 30.72MS/s \\
    FFT size & $N_{FFT}$ & 2048 \\
    \# of used subcarriers & $N_{sc}^D$ & 1200 \\
    OFDM symbols per slot& $N_{s}$ & 7 \\
    CP Type  & - & Normal \\
    Modulation  & - & BPSK, 4/16/64-QAM \\
    \bottomrule[1.0pt]
    \hline
  \end{tabular}
\end{table}

\begin{table}
\caption{SCM channel model parameters.}\label{tab:channel model parameters}
  \centering
  \begin{tabular}{ll}
    \toprule[1.2pt]
    \bf{Parameter} & \bf{Value} \\
    \hline
    Channel model & SCM \\
    Scenario & suburban\_macro \\
    \# of BS antennas  & 128 \\
    \# of UE antennas & 1 \\
    \# of UEs & 12 \\
    Antenna spacing at BS & $0.5{\lambda}$ \\
    \# of multipath & 6 \\
%    Carrier frequency & 4GHz \\
    Delay sampling interval & $\frac{1}{30.72\times 10^{6}}$ \\
    \bottomrule[1.0pt]
    \hline
  \end{tabular}
\end{table}

\subsection{Link-level Procedure}
\begin{figure}[!t]
\centering
\includegraphics[width=0.5\textwidth]{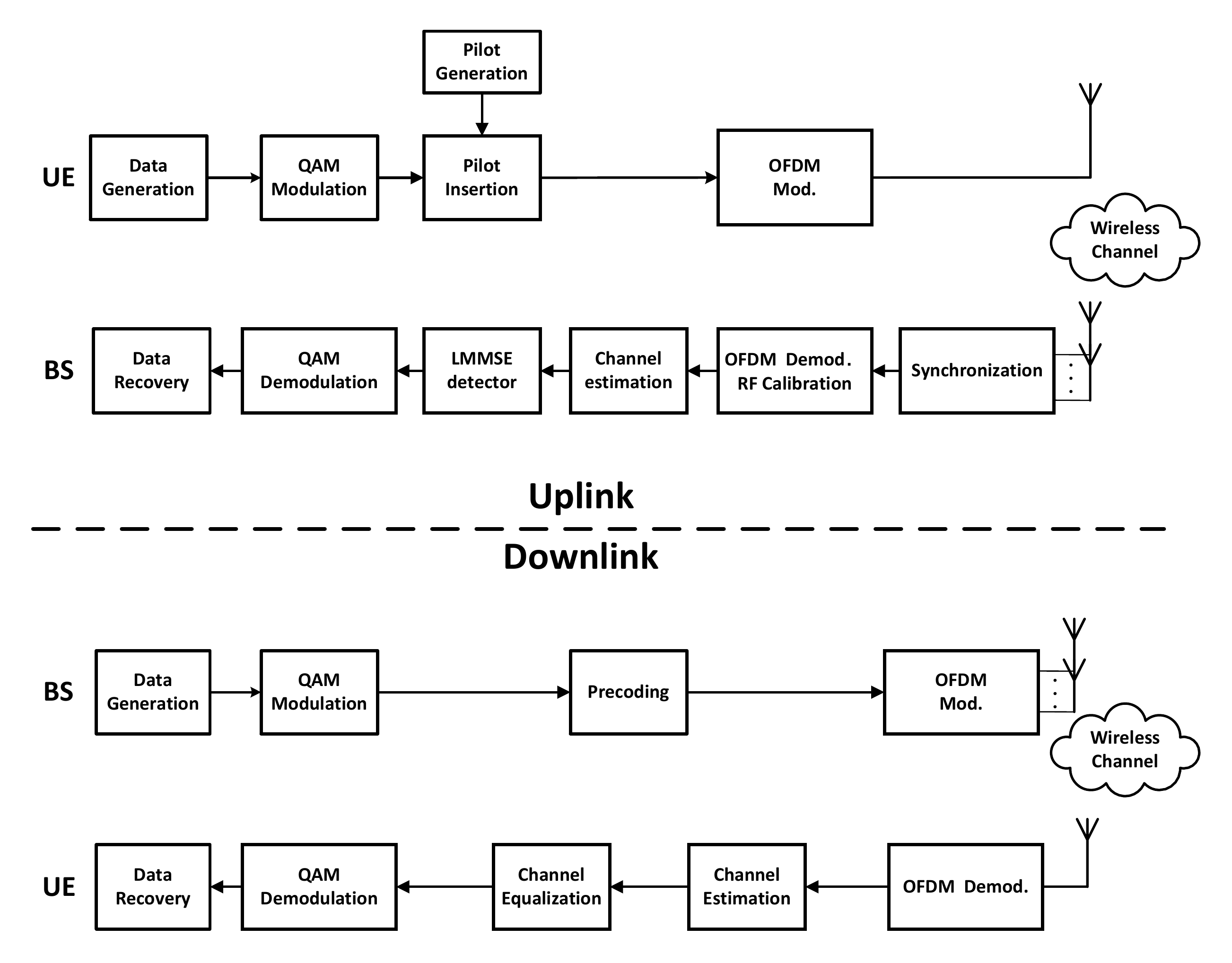}
\caption{The system block diagram of the TDD-based 128 antenna massive MIMO system in link-level simulation. Top: uplink pilot/data transmission. Bottom: downlink pilot/data transmission.}
\label{fig:4}
\end{figure}

According to the system model and simulation parameters given above,
the block diagram of the TDD-based massive MIMO system in link-level simulation is shown in Fig. \ref{fig:4}.

In the uplink, if current OFDM symbol is used for uplink pilot transmission, %according to the frame structure,
then pilot symbols (QPSK modulated) are generated and mapped into resource elements in accordance with the time-frequency resource grids in Fig. \ref{fig:3}.
Otherwise, raw data bits are generated and QAM modulated, after which the acquired data symbols
are mapped into resource blocks in accordance with the time-frequency resource grid.
In the OFDM modulation, IFFT and cyclic prefixing (CP) are carried out.
Either the pilot or data OFDM symbol is then transmitted by the UE through the SCM channel.
At the base station end, synchronization with the UEs is done by PSS.
OFDM demodulation, i.e. FFT and CP removal are then carried out,
followed by reciprocity calibration, LS channel estimation and joint LMMSE detection.
QAM demodulation is then conducted, which recovers the raw data bits and also calculates bit error rate (BER).

For the downlink, it is the inverse process of the uplink.
First of all, raw data bits for the single-antenna users are generated at the BS.
After QAM modulation, precoding (based on the uplink channel estimate) and OFDM modulation,
the users' OFDM modulated signals are transmitted by the massive MIMO base station over the SCM channel.
It is worth noting that we assume the channel is quasi-static within a time slot,
and thus the SCM channel coefficients during one time slot does not change except for the transposing operation between the uplink and the downlink.
At the UE side, similar to the BS in the uplink, OFDM demodulation, LS channel estimation, maximum-ratio combining
and QAM demodulation are conducted in sequential.
%Note that channel coding will be considered in the near future but is not now contained in our link-level simulation, and
Note that there is no specific synchronization in the downlink because the synchronization
has been well achieved in the uplink due to the assumption of time alignment in UEs for simplicity.

Besides, according to the practical measurement, we model the hardware mismatch impairments as complex multiplicative coefficients on subcarriers with unit norm and random phases for both BS and UEs antennas.
Reciprocity calibration is carried out in the initialization of the simulation as later the prototyping system does.
It is worthy to point out that we adopt Pre-precoding Calibration (Pre-Cal) in our link-level simulation,
which is in consistency with our prototyping system design while having little difference with the system model.
As discussed in \cite{rogalin2013hardware}, the reciprocity can be carried out either before or after precoding.
These two scenarios are referred to as Pre-Cal and Post-precoding Calibration (Post-Cal), respectively.
However, the study in \cite{zhang2015large} points out that the Pre-Cal scheme outperforms Post-Cal,
which motivates the use of the Pre-Cal approach in the simulation and in our prototyping system as well.

\subsection{Numerical Results}
In the simulation results presented in the following,
the impacts of reciprocity calibration under different precoding matrices,
the BER for different users with different modulation,
and the throughput of the system are investigated.

\begin{figure}[!t]
\centering
\includegraphics[width=0.45\textwidth]{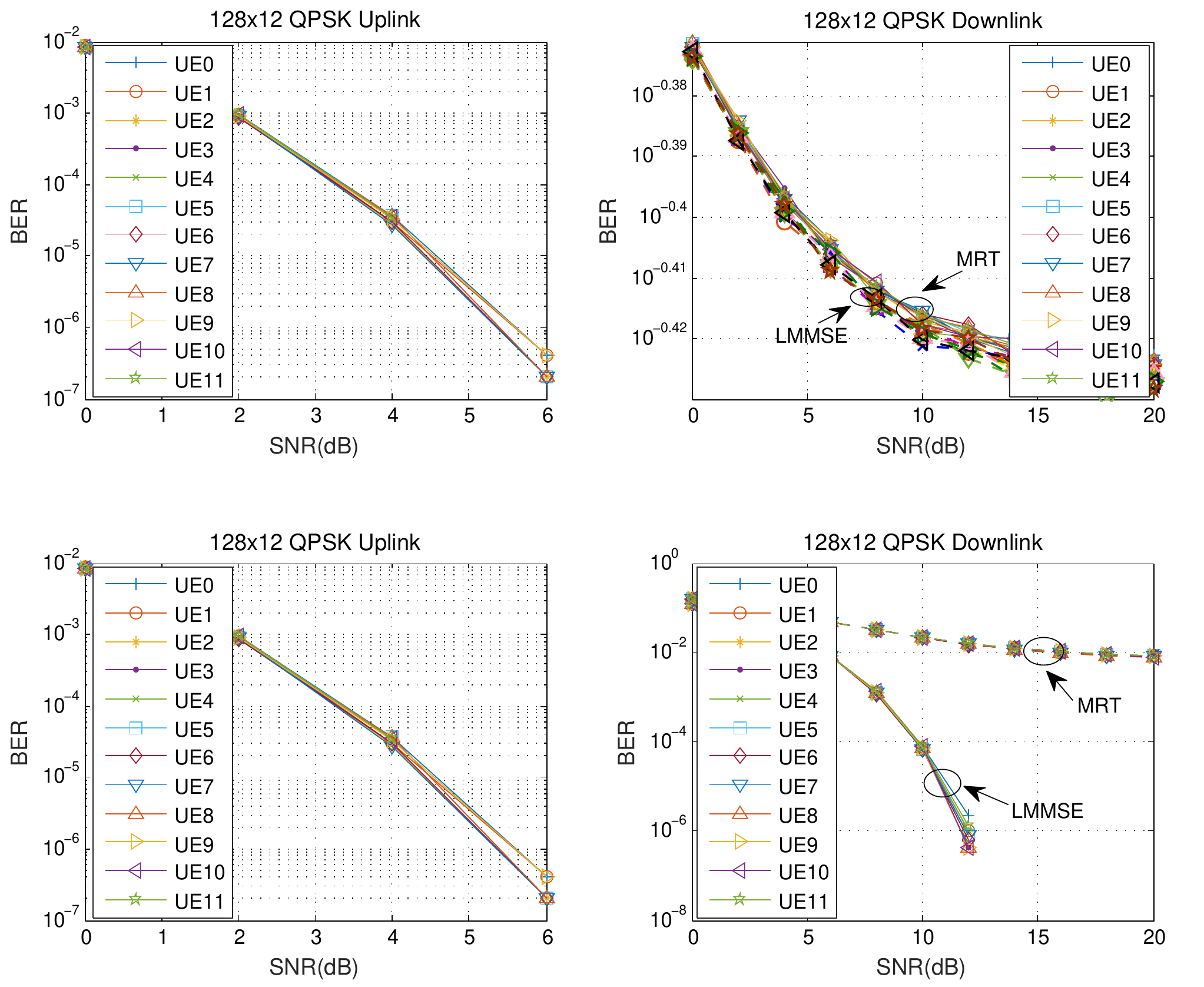}
\caption{The impact of reciprocity calibration under different precoding matrix when $M=128$, $K=12$ and QPSK is used for the $12$ single-antenna users. Top: The BER of uplink(left)/downlink(right) data transmission with reciprocity calibration. Bottom: The BER of uplink(left)/downlink(right) data transmission without reciprocity calibration.}
\label{fig:5}
\end{figure}

\begin{figure}[!t]
\centering
\includegraphics[width=0.45\textwidth]{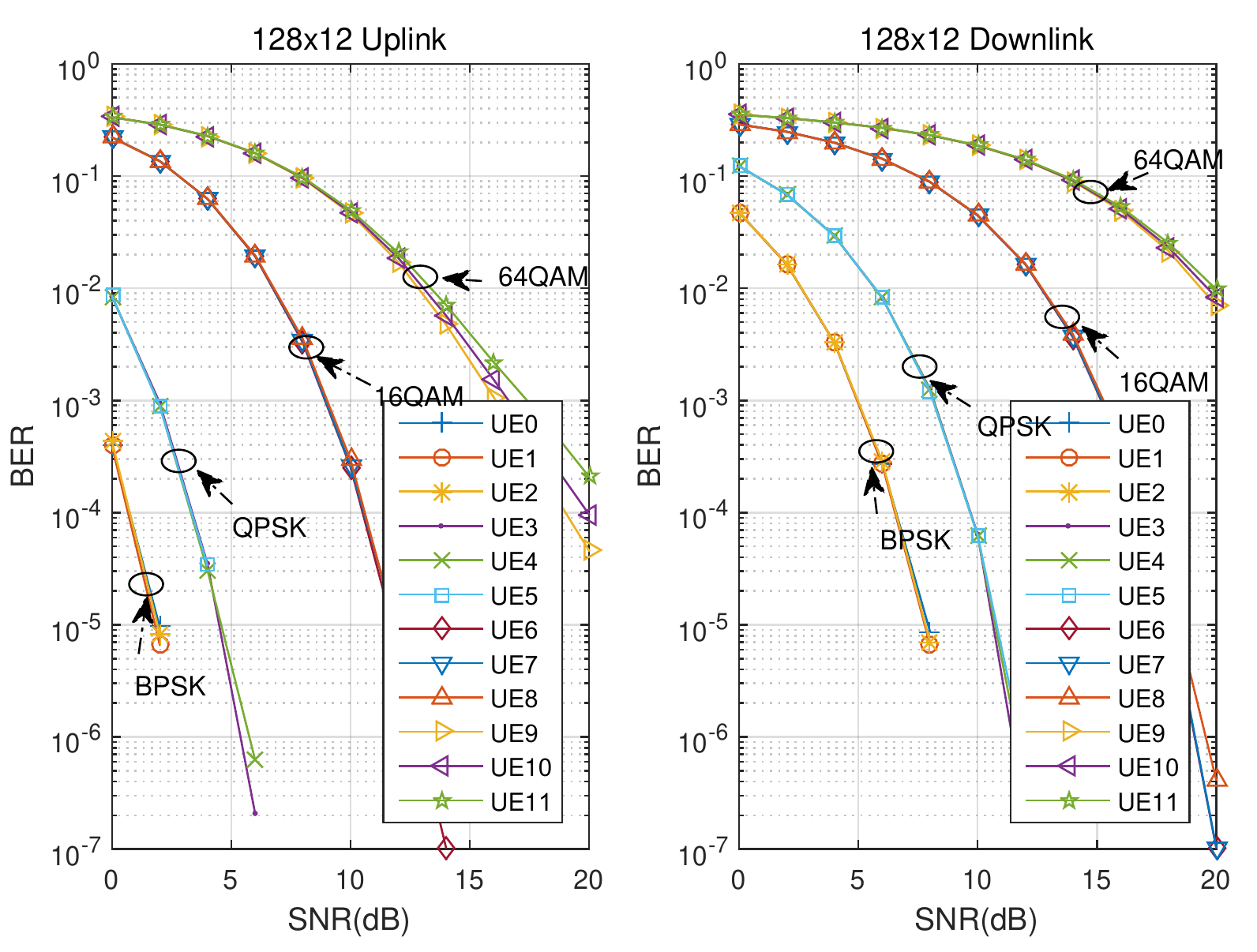}
\caption{The BER for different users with different modulation in uplink and downlink for $M=128$ and $K=12$. BPSK is used for UE0-2, QPSK is used for UE3-5, 16-QAM is used for UE6-8, 64-QAM is used for UE9-11 and reciprocity calibration is also considered. Left: uplink data transmission. Right: downlink data transmission.}
\label{fig:6}
\end{figure}

\begin{figure}[!t]
\centering
\includegraphics[width=0.45\textwidth]{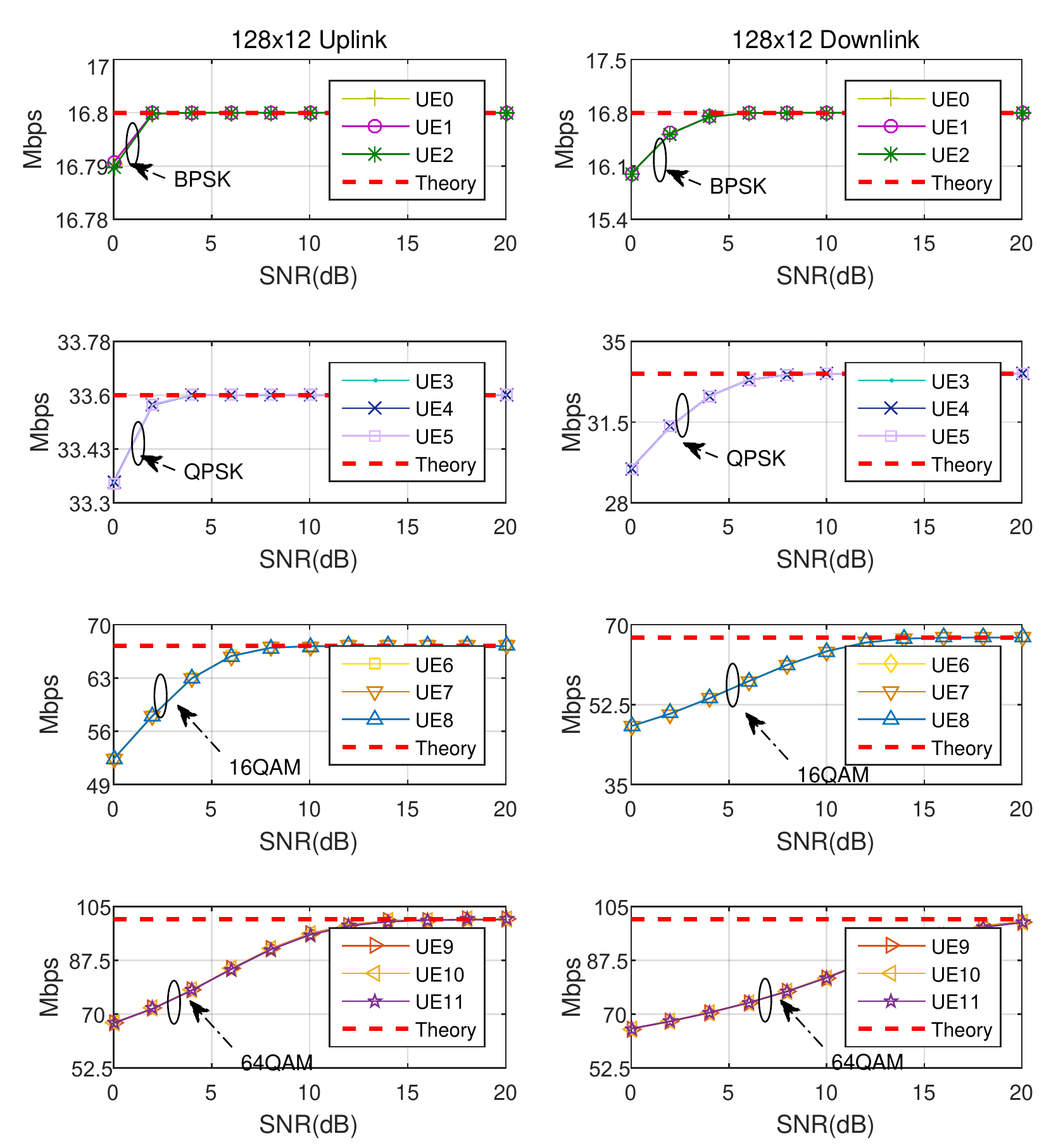}
\caption{Throughput of users when $M=128$ and $K=12$. The theoretical throughput of different QAM modulation under 20MHz bandwidth with OFDM utilized is presented as a baseline for different users, and BPSK is used for UE0-2, QPSK is used for UE3-5, 16-QAM is used for UE6-8, 64-QAM is used for UE9-11. Left: uplink data transmission. Right: downlink data transmission with reciprocity calibration.}
\label{fig:7}
\end{figure}

Fig. \ref{fig:5} shows the impact of reciprocity calibration both in the uplink and the downlink
data transmission under different precoding schemes.
As can be observed, reciprocity calibration has significant impact on downlink data transmission,
but has negligible impact on the uplink.
Regardless of whether or not introducing the reciprocity calibration,
the BER of all the single-antenna users is $10^{-4}$ at $SNR=4dB$ in the uplink.
However, the performance of downlink severely degrades without reciprocity calibration
no matter MRT or LMMSE precoding is employed.
%e.g. $BER=3.8\times 10^{-1}$ at $SNR=20dB$.
This is because the effective uplink channel, which contains the hardware mismatch between
the BS' RX chains and the UEs' TX chains is well estimated by uplink pilot
and the data transmitted from multiuser are jointly processed at BS by making full use of
the estimated effective channel coefficients. %and LMMSE detector.
Nevertheless, the channel reciprocity in TDD mode is destroyed by the hardware mismatch
between BS' TX chains and BS' RX chains.
The precoding matrix constructed from the estimated effective channel cannot effectively
inhibit the interference in the downlink which results in degraded performance.
In addition, with reciprocity calibration, LMMSE precoding outperforms MRT in downlink data transmission.

Fig. \ref{fig:6} and Fig. \ref{fig:7} show the BER and throughput for different users for different modulation schemes.
In Fig. \ref{fig:6}, the uplink data transmission outperforms the downlink due to the joint processing at the BS.
By comparing with Fig. \ref{fig:7}, it is observed that higher modulation order results in worse BER performance but higher throughput as well.
Consequently, there is a tradeoff between system throughput and BER performance.
In our prototyping system, in order to acquire better BER to support video streaming application
in the absence of channel coding, we choose QPSK for all users in the uplink,
which can achieve $268.8Mbps$ peak rate for twelve users.
As for 64-QAM, up to $1.2Gbps$ peak rate can be achieved over a 20MHz bandwidth for twelve users at high SNR.

\begin{figure*}[htb]
\centering
\includegraphics[width=1\textwidth]{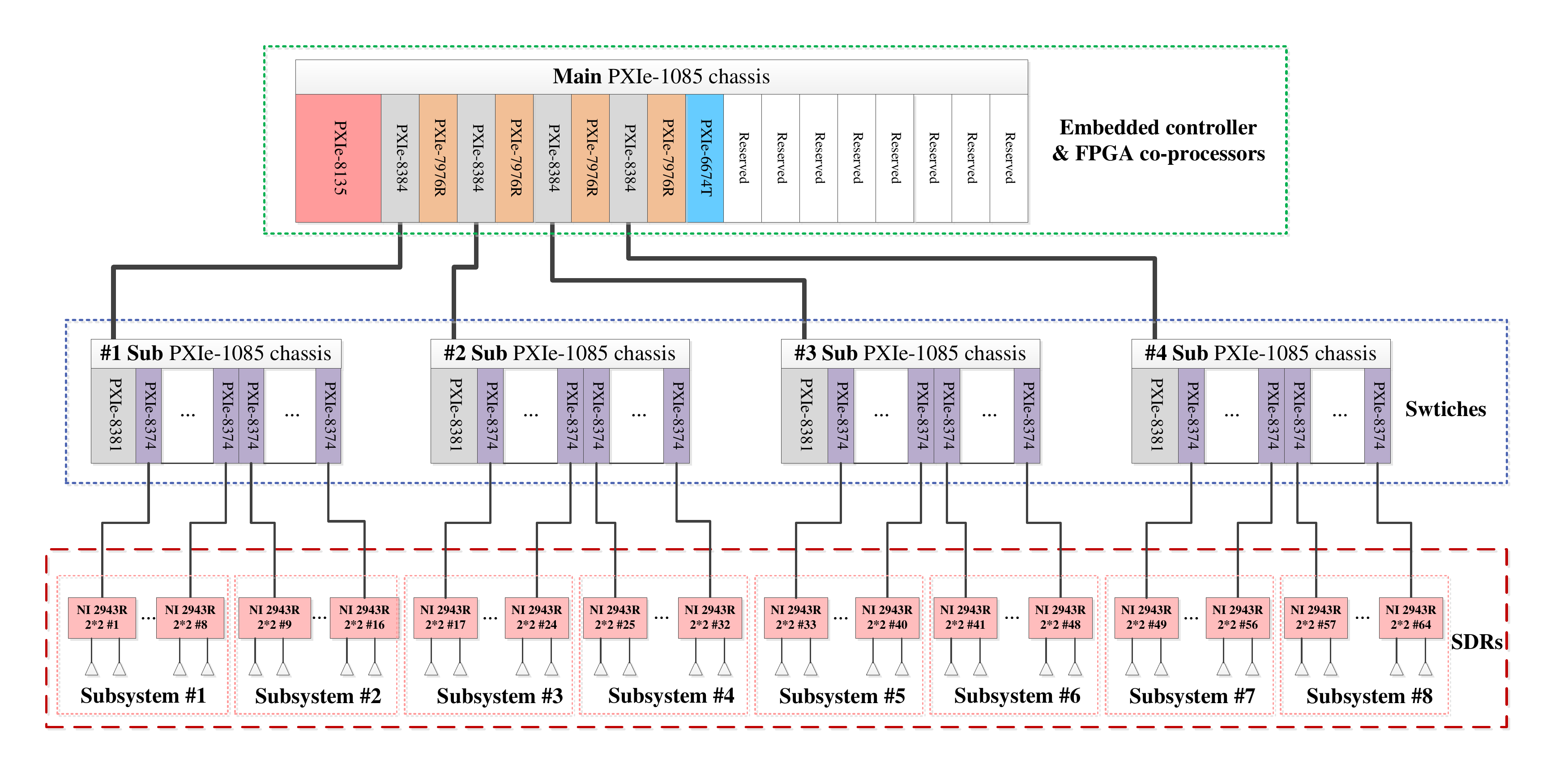}
\caption{The system architecture of our TDD-based 128 antenna massive MIMO prototyping system, the entire system framework is built up of PXIe-1085 chassis in a hierarchical design, where PXIe-1085 chassis serve as switches, data collected by NI 2943Rs will converge at each sub PXIe-1085 chassis, and the main PXIe-1085 chassis is equipped with both PXIe-8135 high-performance embedded controller and PXIe-7976 FPGA co-processor to enhance the data processing capability.}
\label{fig:15}
\end{figure*}

%%%%%%%%%%%%%%%%%%%%%%%%%%%%%%%%%%%%%%%%%%%%%%%%%%%%%%%%%%%%%%%%%%%%%%%%%%%%%%%%%%%%%%%%%%%%%
\section{System Design and Experiment Setup}
In this section, we present the hardware design of our TDD-based 128 antenna massive MIMO prototyping system
including system architecture, the description of major components and experiment setup.
Uplink and downlink data transmission procedures along with hardware devices are also discussed in detail.

\subsection{System Architecture and Experiment Deployment}

\subsubsection{Overview of the system architecture}
The system architecture of our TDD-based 128 antenna massive MIMO prototyping system
based on software defined radio platform (i.e., USRP-RIO manufactured by National Instruments)
combining the clock distribution module and the high data throughput PXI system is showed in Fig. \ref{fig:15}.

A brief introduction of all the hardware components involved in the system block diagram in Fig. \ref{fig:15}
is given in the following.
\begin{itemize}
  \item PXIe-1085 chassis: 3U PXI Express chassis with 18 slots, including 16 hybrid slots and one PXI Express system timing slot. Each hybrid slot has a bandwidth of 4 GB/s and can be connected with an NI 2943R through PXIe-8374.
  \item PXIe-8135: NI PXIe-8135 is a high-performance embedded controller based on Intel Core i7-3610QE processor with 2.3 GHz baseband frequency, 3.3 GHz quad-core CPU and dual-channel 1,600 MHz DDR3 memory.
  \item PXIe-8384/PXIe-8381: x8 Gen 2 cabled PCI Express interface suite, used to connect PXI chassis for the purpose of converging data from sub PXIe-1085 chassis to main PXIe-1085 chassis.
  \item PXIe-6674T: Timing and trigger sync module with on-board highly stable 10 MHz OCXO (sensitivity of 50 ppb). This module is used to generate clock signal and enlarge trigger signal, which can be then routed among multiple devices such as PXI chassis and USRP RIOs to realize precise synchronization of timing and trigger signals across the whole system.
  \item PXIe-7976R: DSP-focused Xilinx Kintex-7 FPGA co-processor, used to help CPU process baseband data such as channel estimation and MIMO detector.
  \item PXIe-8374: MXIe x4 Cabled PCIe interface card, can be used to connect NI 2943R and the PXI chassis for data exchange with a real-time data transfer bandwidth up to 200 MHz and the maximum data transfer rate is 800 MB/s.
  \item NI 2943R: SDR nodes of USRP RIO series, consists of a programmable FPGA (Xilinx Kintex-7) and two RF transceivers of 40MHz bandwidth with center frequency to be configured in the range of 1.2-6GHz, the maximum transmitting power is 15 dBm.
\end{itemize}
According to Fig. \ref{fig:15}, the entire system framework is built up of PXIe-1085 chassis in a hierarchical design, where PXIe-1085 chassis serve as switches.
Data collected by USRP RIOs will converge at each sub PXIe-1085 chassis,
which can connect up to 16 USRP-RIO to construct a MIMO of size 32$\times$32.
Then each sub PXIe-1085 chassis will aggregate data to the main PXIe-1085 chassis through PXIe-8384 and PXIe-8381.
The main PXIe-1085 chassis is equipped with not only the PXIe-8135 high-performance embedded controller,
but also the PXIe-7976 FPGA co-processor to enhance the data processing capability.

\begin{figure}[!t]
\centering
\includegraphics[width=0.45\textwidth]{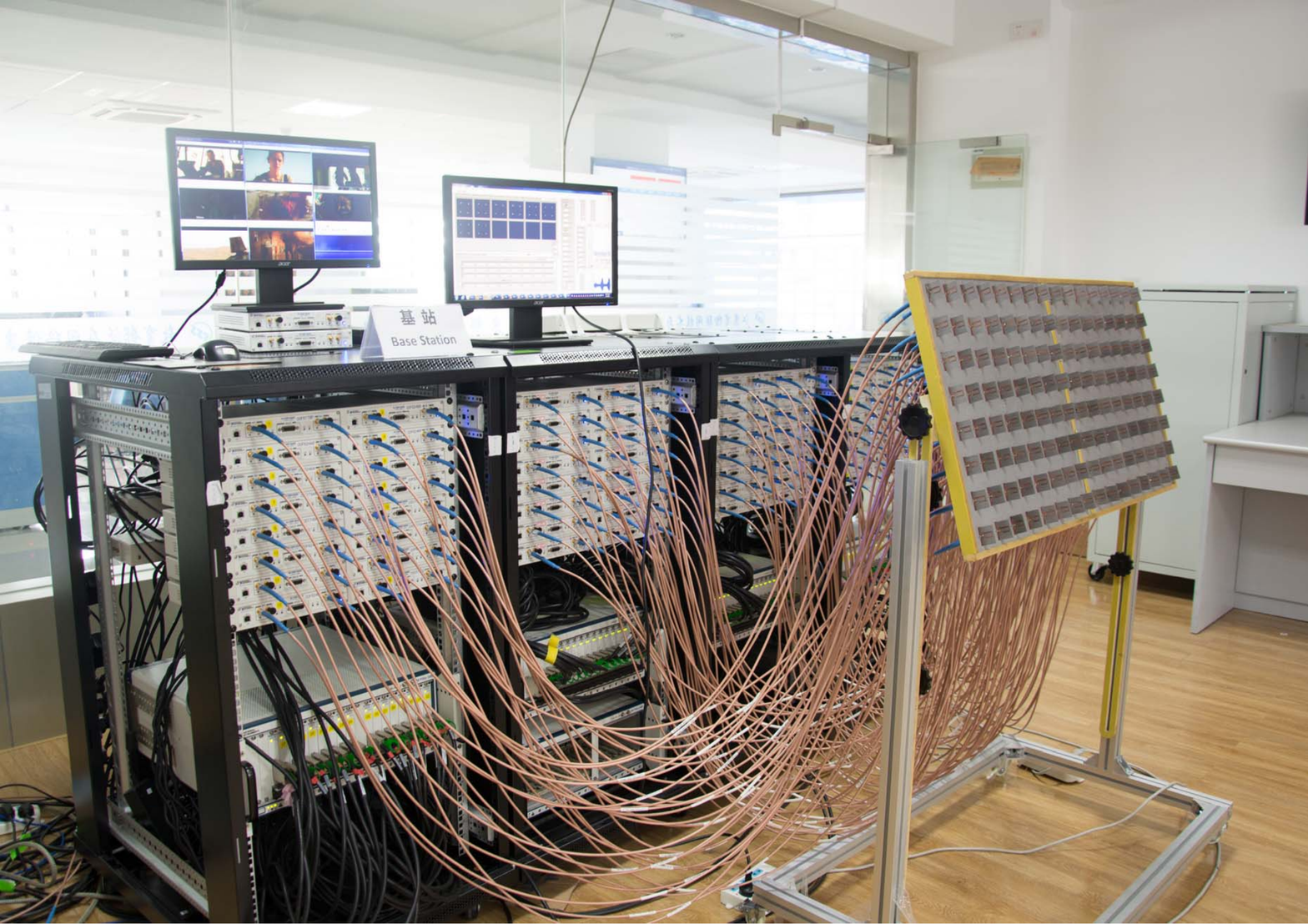}
\caption{Picture of the assembled BS with 8x16 UPA antenna.}
\label{fig:16}
\end{figure}

Besides, as indicated in the diagram, the idea of subsystem is used in our massive MIMO system
to aggregate and transfer baseband data efficiently.
There are a total of eight subsystems in our 128-antenna massive MIMO system with each subsystem
containing eight USRP RIOs, i.e. sixteen antennas.
In each subsystem, one of the eight USRP RIOs serves as data combiner and another serves as data splitter.
All sixteen antennas' whole band (i.e. 20MHz) baseband data will be grouped into consecutive data chunks
where baseband data are aligned with antenna index in data combiner and data splitter,
except the difference that in data combiner, baseband data are aggregated in current subsystem
and will be distributed to sub-band processors for channel estimation and MIMO detection subsequently,
while in data splitter, precoded baseband data are aggregated from sub-band processors and will then be distributed to sixteen antennas in current subsystem. Note that each sub-band processor is related to a sub-band partition, in order to improve system scalability and meet the latency and hardware resource constraints, we partition 20MHz baseband data into four sub-bands. Hence all the eight subsystems will partition their baseband data into four consecutive uniform sub-bands in the unit of data chunk in data combiner respectively. Reversely, all the eight subsystems converge their sixteen antennas' whole-band data from four sub-band processors in data splitter. The embedded controller is responsible for finishing hardware configuration and initialization, displaying the received constellation in uplink and generating raw bits for multi-user in downlink. Picture of the assembled 128 antenna base station is supplied in Fig. \ref{fig:16}.
A 8x16 uniform planar antenna array constituted by dipole element is allocated in front of the rack and connect with NI 2943Rs through SMA cables. Since each NI 2943R has two RF chains, our system needs 64 NI 2943Rs which are divided equally and installed on 4 cabinets. Each cabinet is equipped with 16 NI 2943Rs and one PXIe-1085 chassis making up 2 subsystems except the second one from left which is the main cabinet and consequently equipped with two PXIe-1085 chassis , of which the middle is sub PXIe-1085 chassis and the bottom is main PXIe-1085 chassis.

\subsubsection{Synchronization}
\begin{figure}[!t]
\centering
\includegraphics[width=0.5\textwidth]{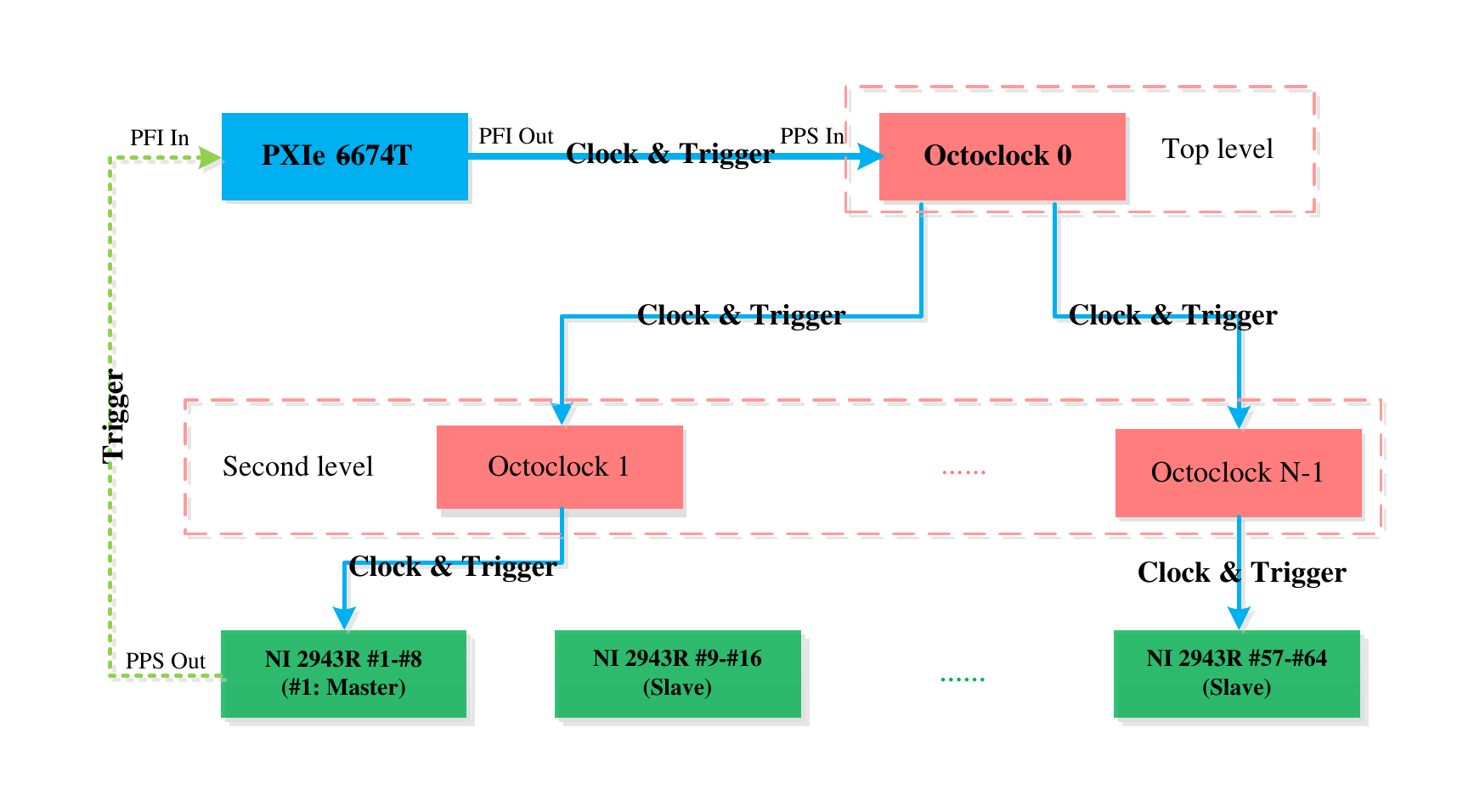}
\caption{Clock and trigger signal distribution network. Firstly, PXIe-6674T generates a 10 MHz reference clock and provides a digital trigger signal to the top-level OctoClock module. Then, the amplified reference clock signal and trigger signal are distributed to eight second-level OctoClock modules to do further amplification and distribution. Finally, each second-level OctoClock module amplifies and distributes reference clock signal and trigger signal to eight USRP RIO devices.}
\label{fig:10}
\end{figure}

Timing and synchronization are critical for multi-device systems especially massive MIMO system that needs the deployment of a large number of radio devices. There are two challenges in timing and synchronization for our massive MIMO system, one is the timing and synchronization among radio devices at BS, the other is timing and synchronization between BS and UEs. In order to solve the problem of timing and synchronization among radio devices at BS, clock and trigger signal distribution network is utilized by the use of OctoClock module. Fig. \ref{fig:10} presents the clock and trigger signal distribution network. The OctoClock module in the diagram is a signal amplifier and distribution module, it can use an external 10 MHz reference clock and the pulse per second (PPS) signal as clock source and trigger signal source. And the input clock signal and trigger signal will be then amplified and distributed to eight channels in OctoClock to provide synchronization of timing and trigger signals for next eight OctoClock modules or eight USRP RIO devices depending on the peripheral connected.

The principle of the clock and trigger signal distribution network adopted in our massive MIMO system can be summarized as following: Firstly, the timing and sync module PXIe-6674T which has an oven-controlled crystal oscillator (OCXO) generates a stable and precise 10 MHz reference clock (sensitivity of 50 ppb) and provides a digital trigger signal to the top-level OctoClock module. Then, the amplified reference clock signal and trigger signal are distributed to eight second-level OctoClock modules to do further amplification and distribution. Finally, each second-level OctoClock module amplifies and distributes reference clock signal and trigger signal to eight USRP RIO devices, therefore, all 128 antennas of the 64 USRP RIOs share the same reference clock signal and trigger signal so that all radio devices at BS can start data collection and generation synchronously.

As for timing and synchronization between BS and UEs, we make use of primary synchronization signal (PSS) like LTE: UEs transmit PSS to BS at first, after receiving PSS, BS performs a cross correlation of the received PSS with the original PSS and finds the peak index among a 10ms radio frame, then the peak index is conveyed to 64 USRP RIOs by the embedded controller at BS to align all the radio devices' time, thus synchronization between BS and UEs is achieved. Note that carrier offset compensation need to be considered due to the sampling clock frequency offset between BS and UEs.

\subsubsection{Reciprocity Calibration}
Given that for all multi-user beamforming techniques using linear precoding, it is sufficient for beamforming antennas to have a relatively accurate channel state information i.e. a constant multiplicative factor across base station antennas do not affect multi-user interference, we realize pre-precoding relative calibration method as mentioned in \cite{shepard2012argos,vieira2014reciprocity} in our prototyping system. The calibration process is showed in Algorithm \ref{alg1:calibration process}.  Note that BS needs to finish RF configuration and initialization before the start of reciprocity calibration process, UEs need to keep silent during the whole process, and make sure there is also no other transmitting signals in the same frequency during calibration.
\begin{algorithm}[htb]
\caption{Relative Reciprocity Calibration Process}
\label{alg1:calibration process}
\begin{algorithmic}[1]
\REQUIRE reciprocity coefficients $b_{i,j,n}$, antenna number $M$, subcarrier number $N$, average reciprocity coefficients $b_{i,j}$, $i=1 \dots M, j=1 \dots M, n=1 \dots N$.\\
\STATE \textbf{Initialization}: set $b_{i,j,n}=1$ for all $i,j,n$. \\
\STATE \textbf{for} {$i = 1:1:M$}
\STATE \quad antenna $i$ transmits reference signal
\STATE \quad all $M$ antennas receive and process the signal
\STATE \quad note down $b_{i,j,n}$, for $j=1 \dots M, n=1 \dots N$.
\STATE \textbf{end for}
\STATE select one of the $M$ antennas as reference antenna $m_{ref}$.
\STATE \textbf{for} {$i = 1:1:M, i \ne m_{ref}$}
\STATE \quad \textbf{for} {$n = 1:1:N$}
\STATE \quad \quad $b_{i,m_{ref},n}=\frac{1}{N}\sum\limits_{n = 1}^N {{b_{i,{m_{ref}},n}}} $
\STATE \quad \quad $b_{i,m_{ref},n}=b_{i,m_{ref},n} / \left\| {{b_{i,{m_{ref}},n}}} \right\| $
\STATE \quad \textbf{end for}
\STATE \textbf{end for}
\ENSURE $b_{i,m_{ref},n}$ for $i=1 \dots M, i \ne m_{ref}, n=1 \dots N$.
\end{algorithmic}
\end{algorithm}

\subsubsection{Antenna Array}
\begin{figure}[!t]
\centering
\includegraphics[width=0.5\textwidth]{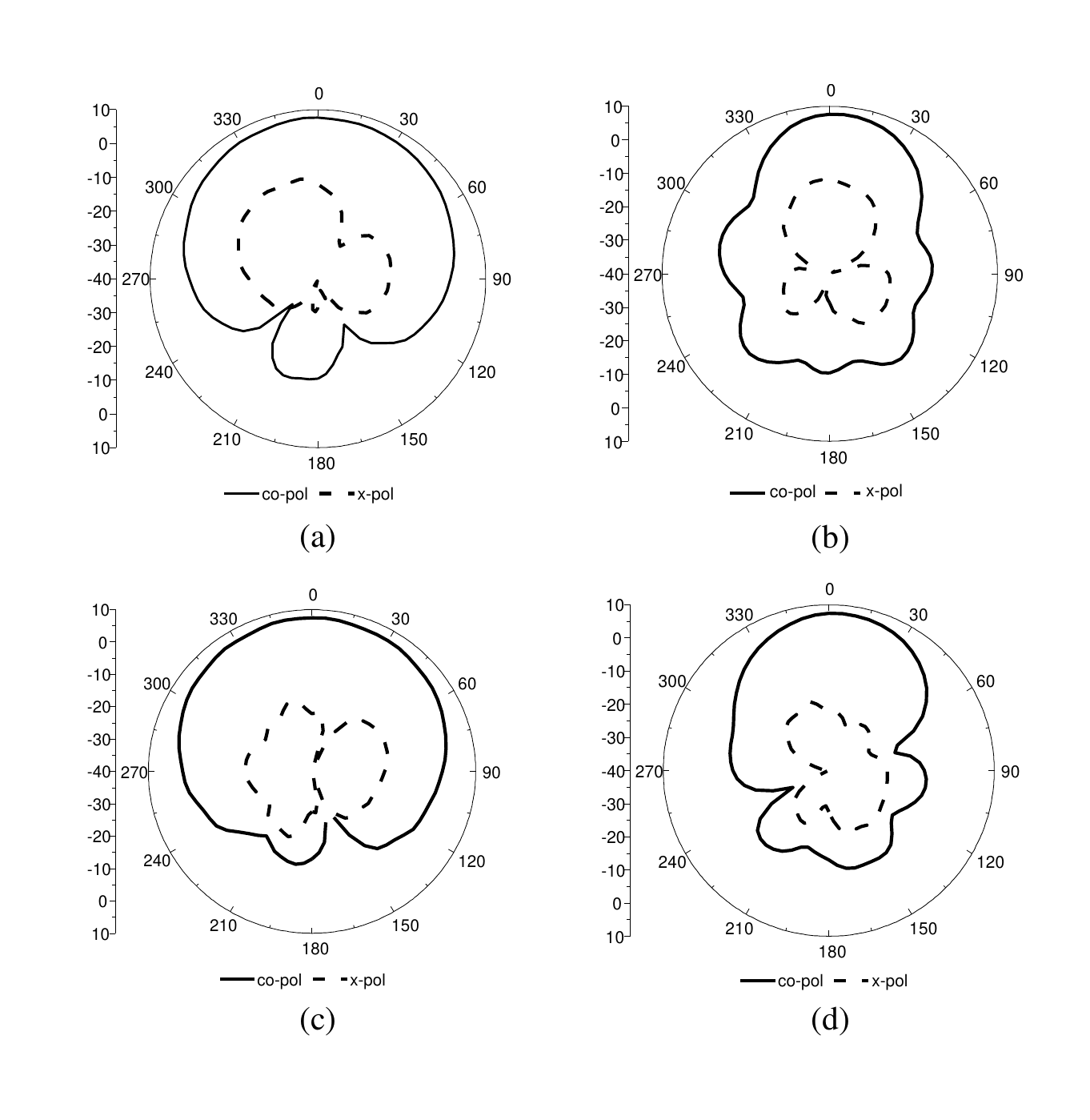}
\caption{Radiation patterns of principal planes, H plane is parallel to the ground and E plane is perpendicular to the ground, (a) H plane @3.8GHz, (b) E plane @3.8GHz, (c) H plane @4.1GHz, (d) E plane @4.1GHz.}
\label{fig:8}
\end{figure}

\begin{figure}[!t]
\centering
\includegraphics[width=0.5\textwidth]{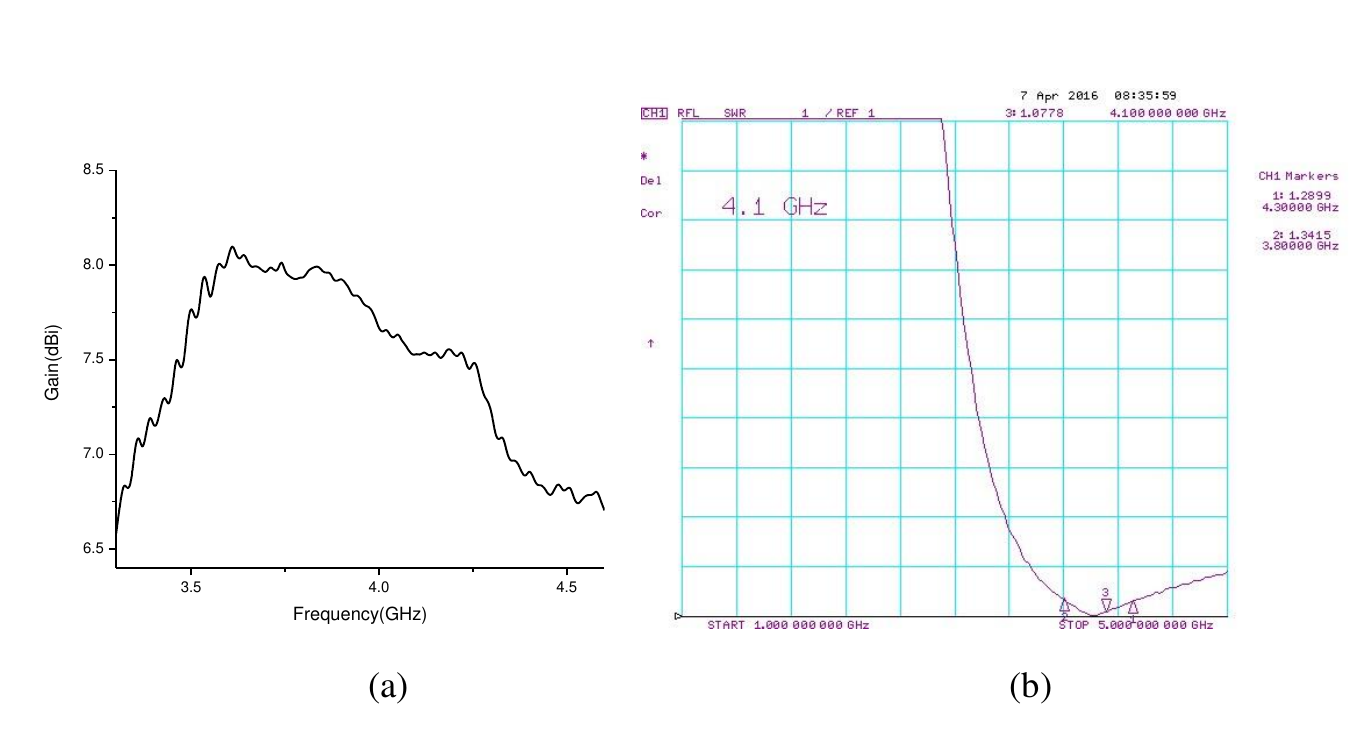}
\caption{Measured gain and SWR frequency response characteristics of the antenna element, (a) gain, (b) standing wave ratio (SWR).}
\label{fig:9}
\end{figure}

A 128-element uniform planar array (UPA) is designed to serve as the base station antenna array of our massive MIMO prototype system.
The array is composed of eight 16-element, linear sub-arrays.
All elements are printed dipoles mounted above metallic reflectors, and they are operating at 3.8-4.3GHz with a uniform separation of 0.8 wavelength.
The antenna array can be employed to verify either the beamforming algorithms or the three-dimensional (3D) MIMO ones.
Measured performance of the antenna element is tabulated in Table \ref{tab:antenna parameters}. The dipole element is measured by using Agilent＊s 8720ET vector network analyzer (VNA) and Microwave Vision＊s Starlab near field antenna measurement system. As can be observed from Fig. \ref{fig:8}, Fig. \ref{fig:9} and Table \ref{tab:antenna parameters}, the SWR of the dipole is lower than 1.4 from 3.8-4.3GHz. The dipole exhibits a stable unidirectional, linearly polarized radiation pattern within its impedance bandwidth: The half-power beam width of the E- and the H-plane is $55^{\circ}$ and $100^{\circ}$, respectively. The front-to-back ratio of the antenna is higher than 22dB and the in-band average gain is 7.7dBi.
\begin{table}
\caption{Measured antenna performance.}\label{tab:antenna parameters}
  \centering
  \begin{tabular}{ll}
    \toprule[1.2pt]
    \bf{Parameter} & \bf{Performance} \\
    \hline
    \multirow{3}*{Standing wave ratio (SWR)} & $<$1.1@ 3.9-4.1GHz, \\
                                             & $<$1.4 @ 3.8-4.3GHz,\\
                                             & and $<$2 @ 3.5-4.5GHz.\\
    Element isolation & $>$25dB \\
    Average gain (dBi)  & 7.7dBi \\
    Polarization & Vertically linear polarization \\
    Launcher & 3.5mm SMA-F \\
    \bottomrule[1.0pt]
    \hline
  \end{tabular}
\end{table}

\subsubsection{User Equipment} 
\begin{figure}[!t]
\centering
\includegraphics[width=0.45\textwidth]{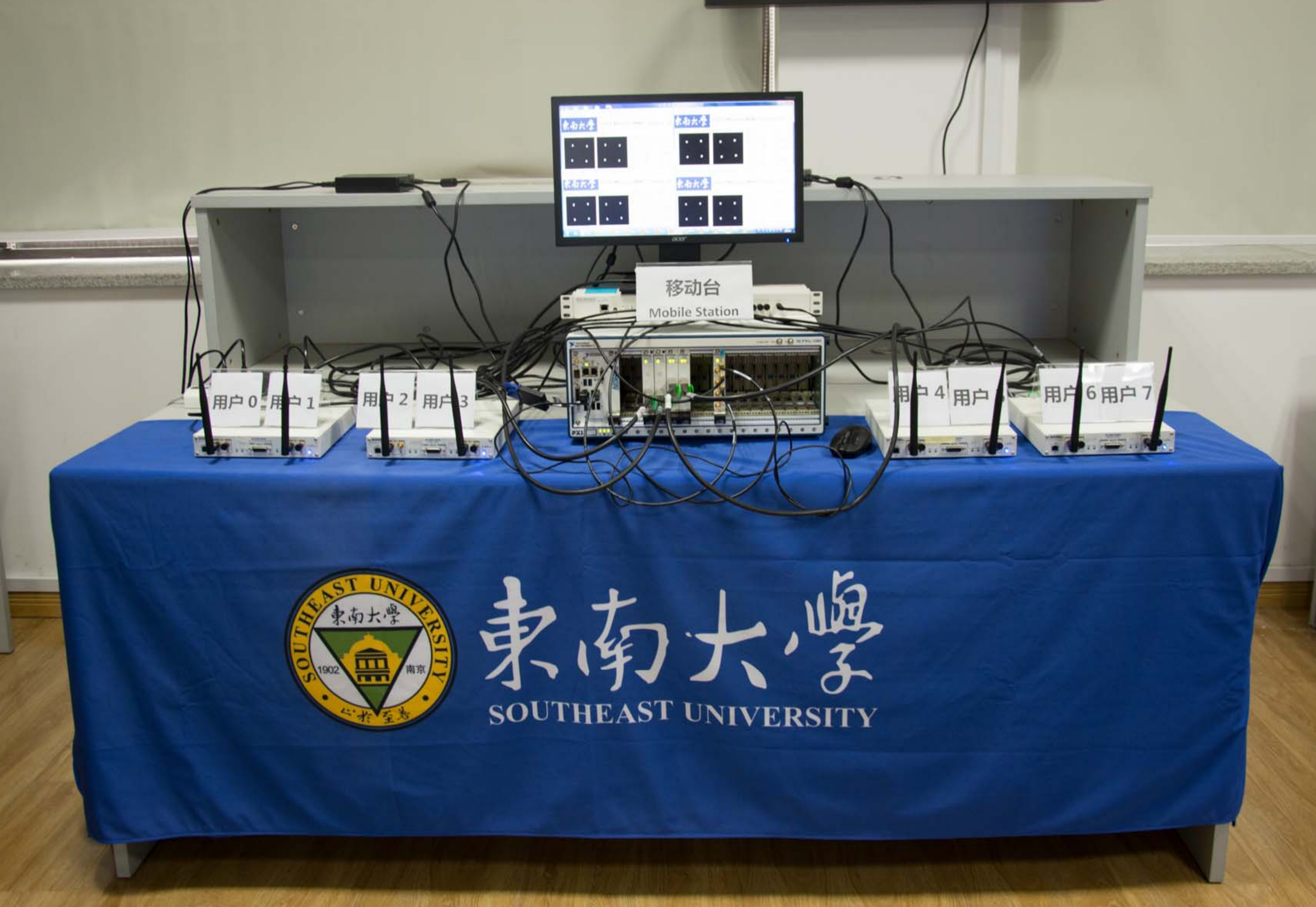}
\caption{User Equipments (UEs): eight single-antenna users.}
\label{fig:12}
\end{figure}

Four SDRs (NI 2943Rs) are used at the terminal ends to emulate eight single-antenna users as shown in Fig. \ref{fig:12}. To simplify the hardware implementation of synchronization between BS and multiple single-antenna users, 10MHz reference clock signal is shared among the four SDRs. The details of hardware implementation for each single-antenna user is provided in Fig. \ref{fig:4}, where data generation/recovery is implemented in embedded controller or computer and the rests are programmed in FPGA contained in SDRs.

\subsubsection{Experiment Deployment}
\begin{figure}[!t]
\centering
\includegraphics[width=0.5\textwidth]{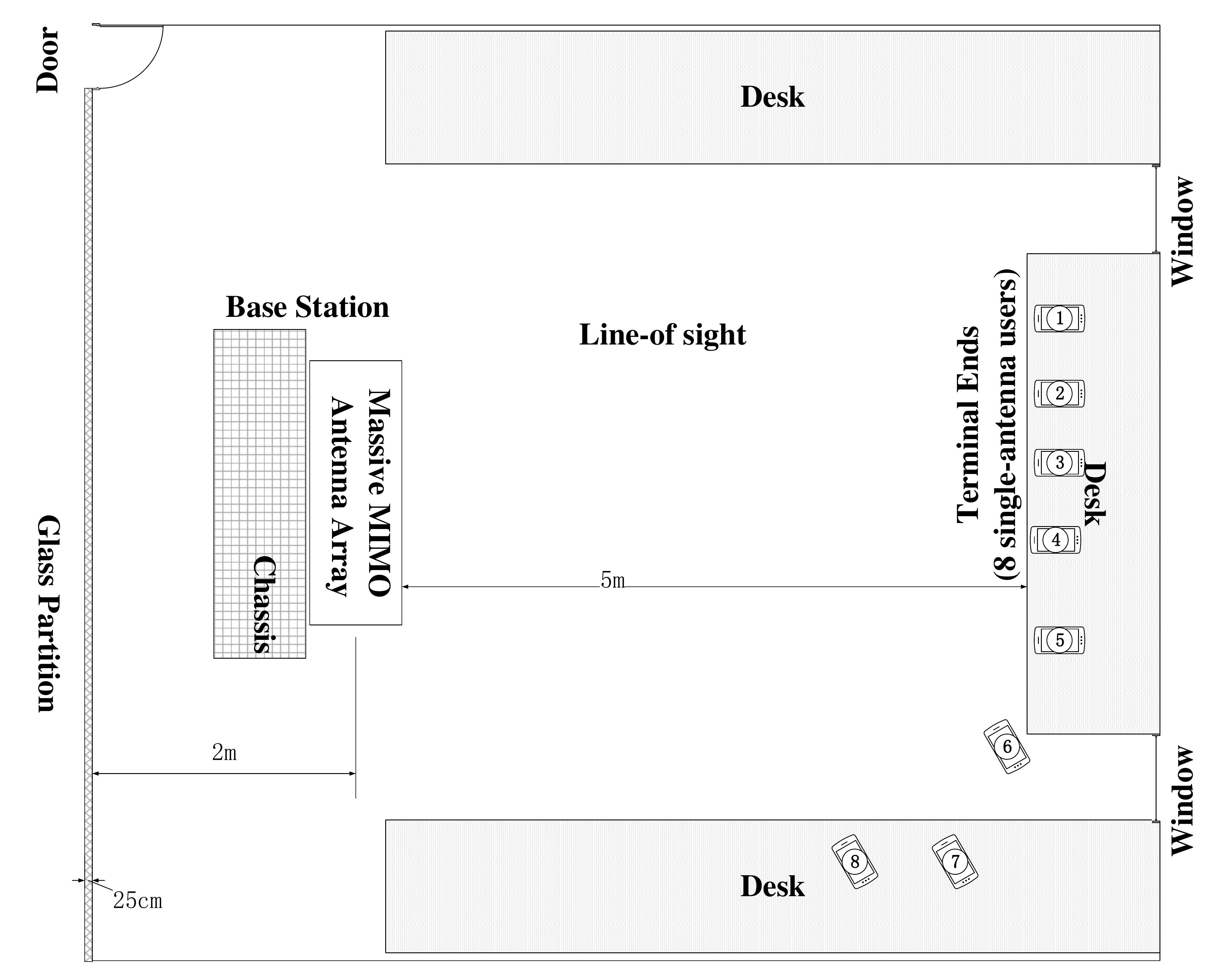}
\caption{The measured environment and experiment deployment.}
\label{fig:11}
\end{figure}

The experiments are conducted in a typical indoor office environment and its deployment is offered in Fig. \ref{fig:11}. The 128-element UPA with 1.2m height is fixed near the chassis, and the eight horn antennas related to eight single-antenna users is placed at eight line-of-sight (LOS) points marked with $1,2, \dots 8$. A series of experiments are carried out in the deployment including multi-user massive MIMO channel measurement, multiple video streaming transmission in uplink, multi-user beamforming data transmission in downlink and the performance of the relative reciprocity calibration method. Experiment results are illustrated in Section V.

\subsection{Uplink Data Transmission Procedure}
\begin{figure}[!t]
\centering
\includegraphics[width=0.5\textwidth]{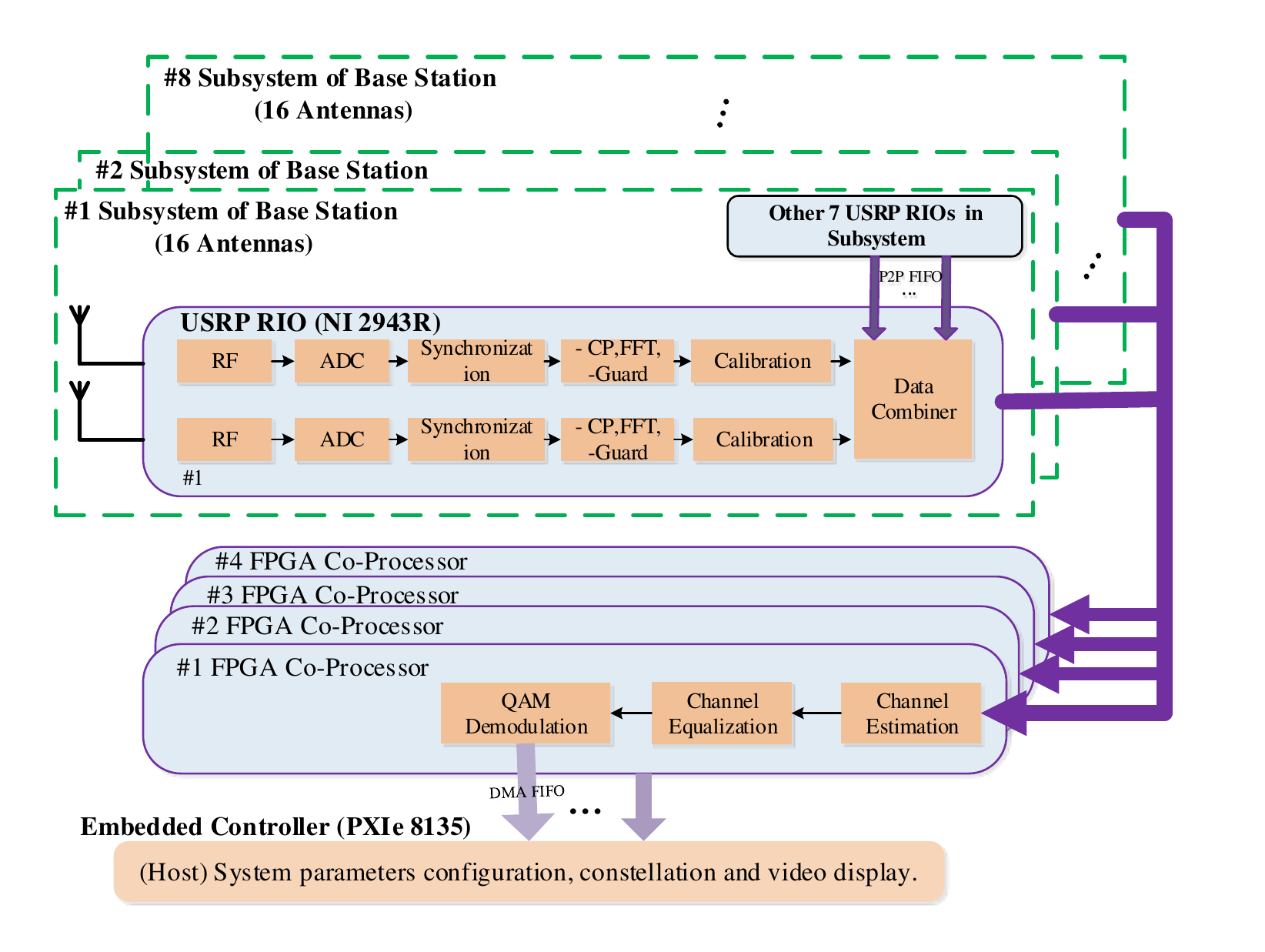}
\caption{System block diagram related to hardware implementation for uplink data transmission at base station.}
\label{fig:13}
\end{figure}

Fig. \ref{fig:13} presents the system block diagram related to hardware implementation for uplink. As is shown in the figure, for uplink, the RF signals acquired by 64 NI 2943Rs, i.e. 128 antennas, firstly go through the 128 RF chains and perform low noise amplification, down conversion and ADC sampling and quantization, and then the high rate samples from ADC are sent to each NI 2943Rs' FPGA for IQ imbalance correction, frequency shift correction, digital down sampling, OFDM demodulation and reciprocity calibration, after that these obtained valid baseband data are aggregated and distributed to four FPGA co-processors for further baseband processing by data combiners through switches (e.g. PXIe-1085) in each subsystem. Finally, these recovered data are conveyed to the embedded controller by co-processors for further analysis and display. In our prototyping system, the conversion accuracy of ADC is 12 bit, thus available data throughput per RF chain is 50.4 MB/s (including I and Q). Each subsystem contains 16 RF chains, therefore available data throughput per subsystem is 806.4 MB/s. There are total 8 subsystems connected with the main switch thus the available data throughput in main switch will be about 6.5 GB/s.

\subsection{Downlink Data Transmission Procedure}
\begin{figure}[htb]
\centering
\includegraphics[width=0.5\textwidth]{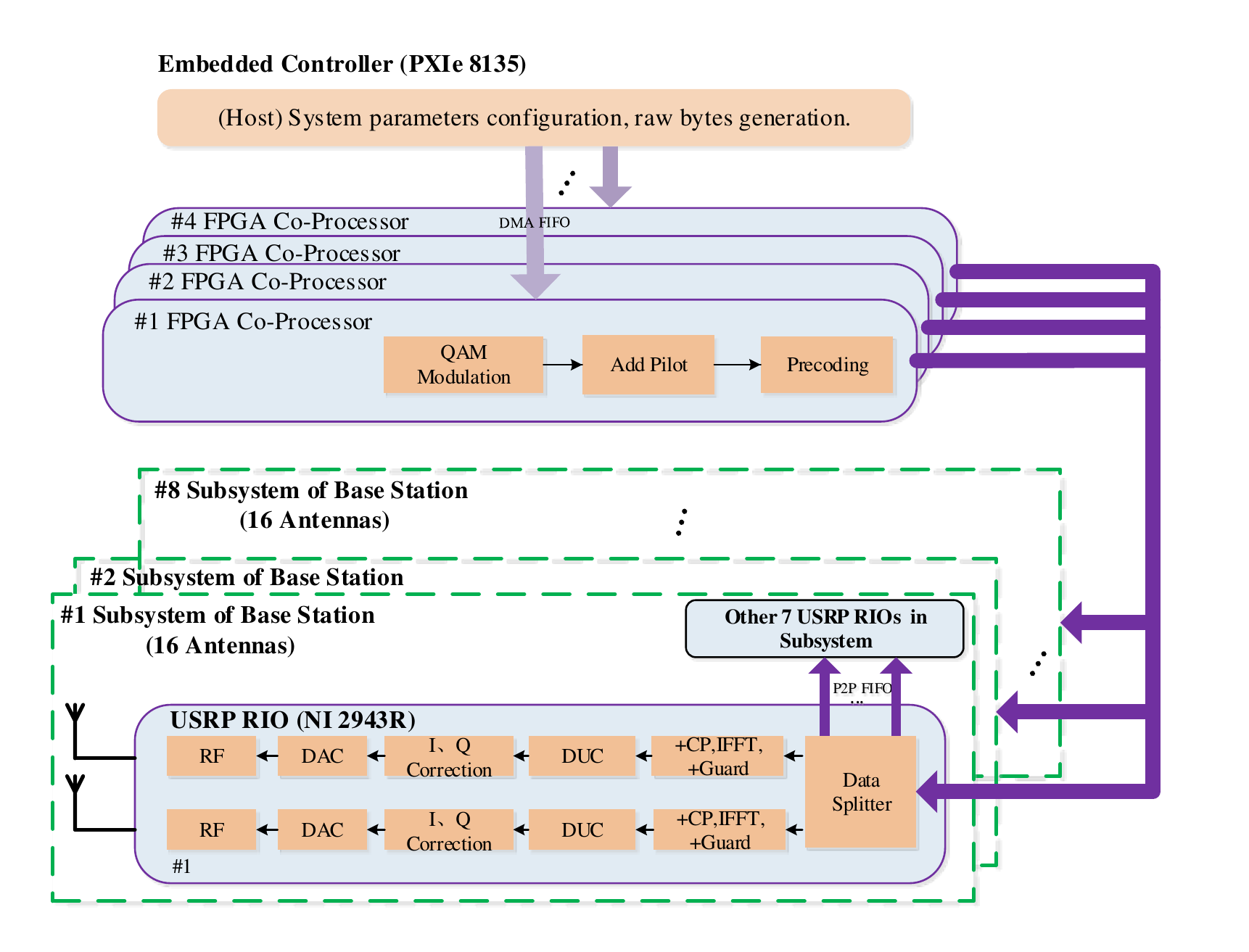}
\caption{System block diagram related to hardware implementation for downlink data transmission at base station.}
\label{fig:14}
\end{figure}

Downlink data transmission is a reverse process compared with uplink, the system block diagram related to hardware implementation for downlink is displayed in Fig. \ref{fig:14}. For downlink, raw data bytes generated by the embedded controller are firstly transferred to four co-processors for precoding, and then these precoded data are aggregated and distributed to each NI 2943R by data splitters through switches (e.g. PXIe-1085) in each subsystem. After OFDM modulation, digital up sampling, frequency shift correction and IQ imbalance correction in the FPGA of each NI 2943R, the high rate data bytes will be conveyed to each RF chain for digital to analog conversion and up conversion, and be sent to the air by antennas finally.

%%%%%%%%%%%%%%%%%%%%%%%%%%%%%%%%%%%%%%%%%%%%%%%%%%%%%%%%%%%%%%%%%%%%%%%%%%%%%%%%%%%%%%%%%%%%%
\section{Experiment Results}
By running the TDD-based 128-antenna massive MIMO prototyping system, experiments are carried out to test performance of our design.
These experiments include multi-user massive MIMO channel measurement, multiple video streaming transmission in the uplink, multi-user beamforming data transmission in the downlink and the performance of the relative reciprocity calibration method.
Experiment results are presented and discussed in this section.

\begin{figure}[htb]
\centering
\includegraphics[width=0.5\textwidth]{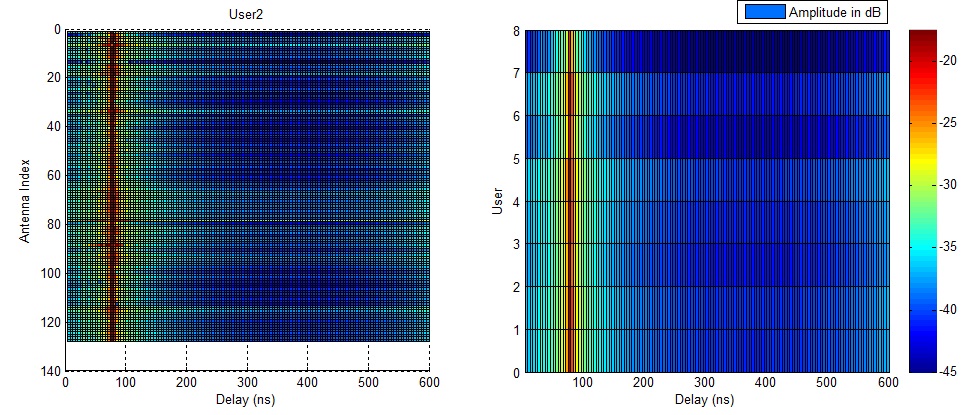}
\caption{Left: time-domain impulse response of uplink channel for user2, the horizontal axis is delay (ns) and the vertical axis is antenna index, 128 antennas are configured in BS. Right: time-domain impulse responses of uplink channel for eight single-antenna users, averaged on 128 antennas.}
\label{fig:23}
\end{figure}

\begin{figure}[htb]
\centering
\includegraphics[width=0.45\textwidth]{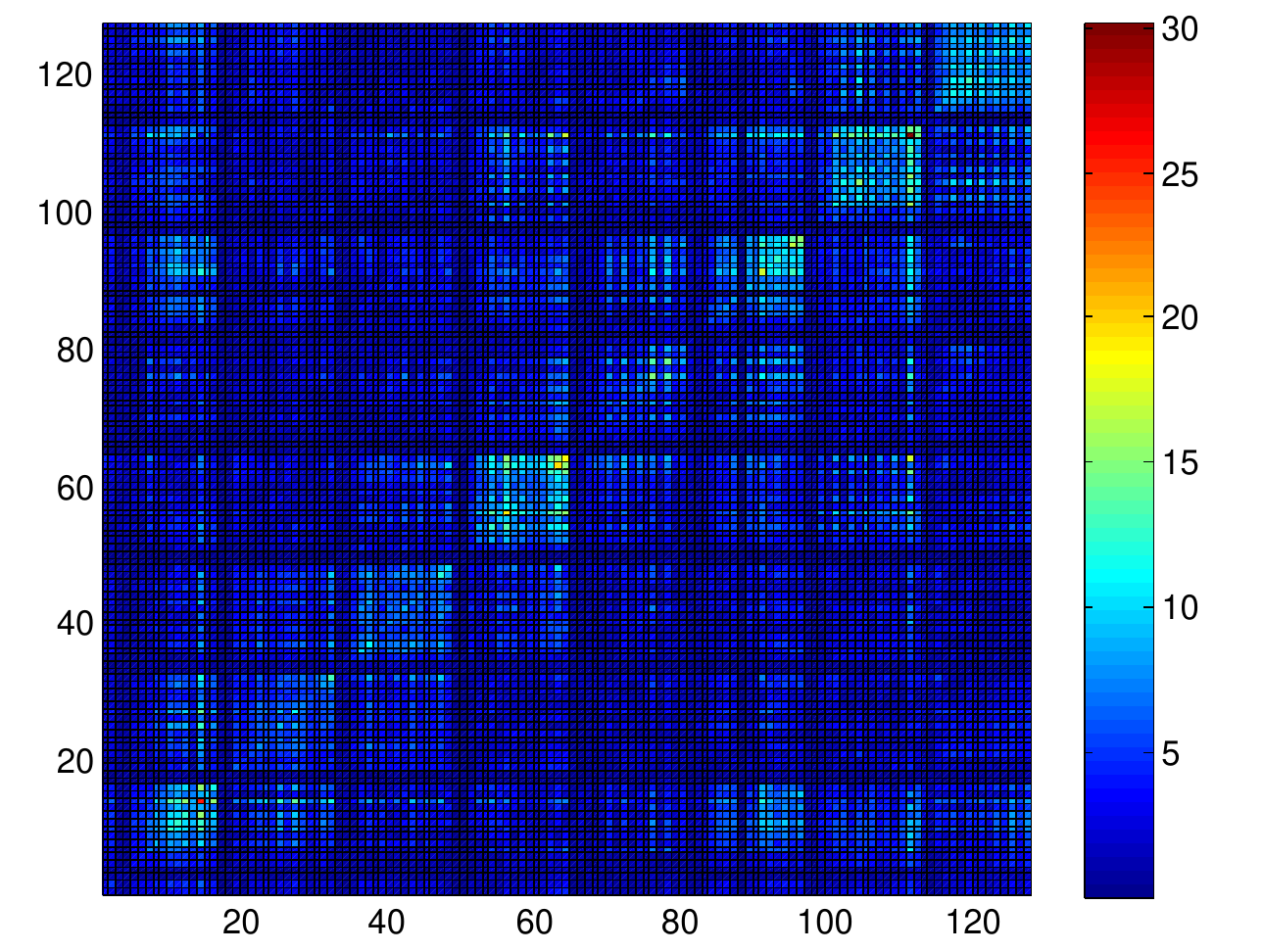}
\caption{Channel correlation matrix on the BS side.}
\label{fig:21}
\end{figure}

\begin{figure}[htb]
\centering
\includegraphics[width=0.4\textwidth]{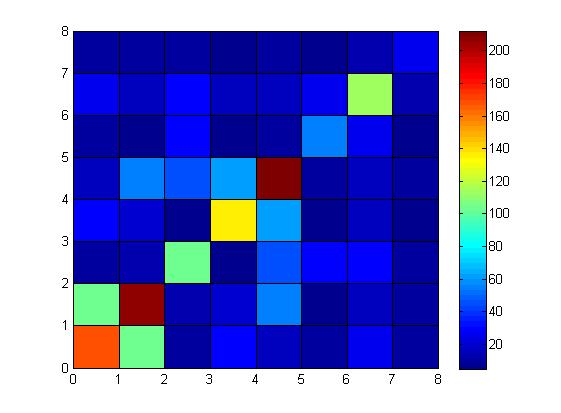}
\caption{Channel correlation matrix on the user side.}
\label{fig:22}
\end{figure}

In order to measure the multi-user massive MIMO channel, pilots orthogonal in the frequency domain
are transmitted by 8 single-antenna users over the same time-frequency resource block
after synchronization between the BS and the UEs using a primary synchronization sequence.
After receiving pilot signals, the BS estimates each user's channel matrix using LS channel estimation with local pilot sequence.
Then the measured channel matrices are further processed and analyzed to obtain results including
channel time-domain impulse response, channel correlation matrix on the BS side
and channel correlation matrix on the user side, which are demonstrated in Figs. \ref{fig:23} through \ref{fig:22}.
Fig. \ref{fig:23} shows that for user2 there is a distinctive planar wavefront with about 33 ns delay spread
despite the little difference among different antennas, which is the same as the other seven users' implied by the averaged impulse response.
Combined with the sample rate of 30.72 MS/s (i.e. 33 ns) for the 20 MHz bandwidth,
the frequency selectivity of the channel is not severe in current deployment.
In addition, the distinctive planar wavefront of the right hand side plot of Fig. \ref{fig:23}
also verifies that the eight users are well time aligned in the uplink.
Fig. \ref{fig:21} and Fig. \ref{fig:22} show the channel correlation matrix.
For channel correlation matrix on the BS side, signal strength is not concentrated on the diagonal line
but on the border of squares.
This is consistent with the geometry of the antenna array used in our prototyping system.
As for the UEs, signal strength is concentrated on the diagonal line as expected,
which indicates that the single-antenna users are independent.

\begin{figure}[htb]
\centering
\includegraphics[width=0.45\textwidth]{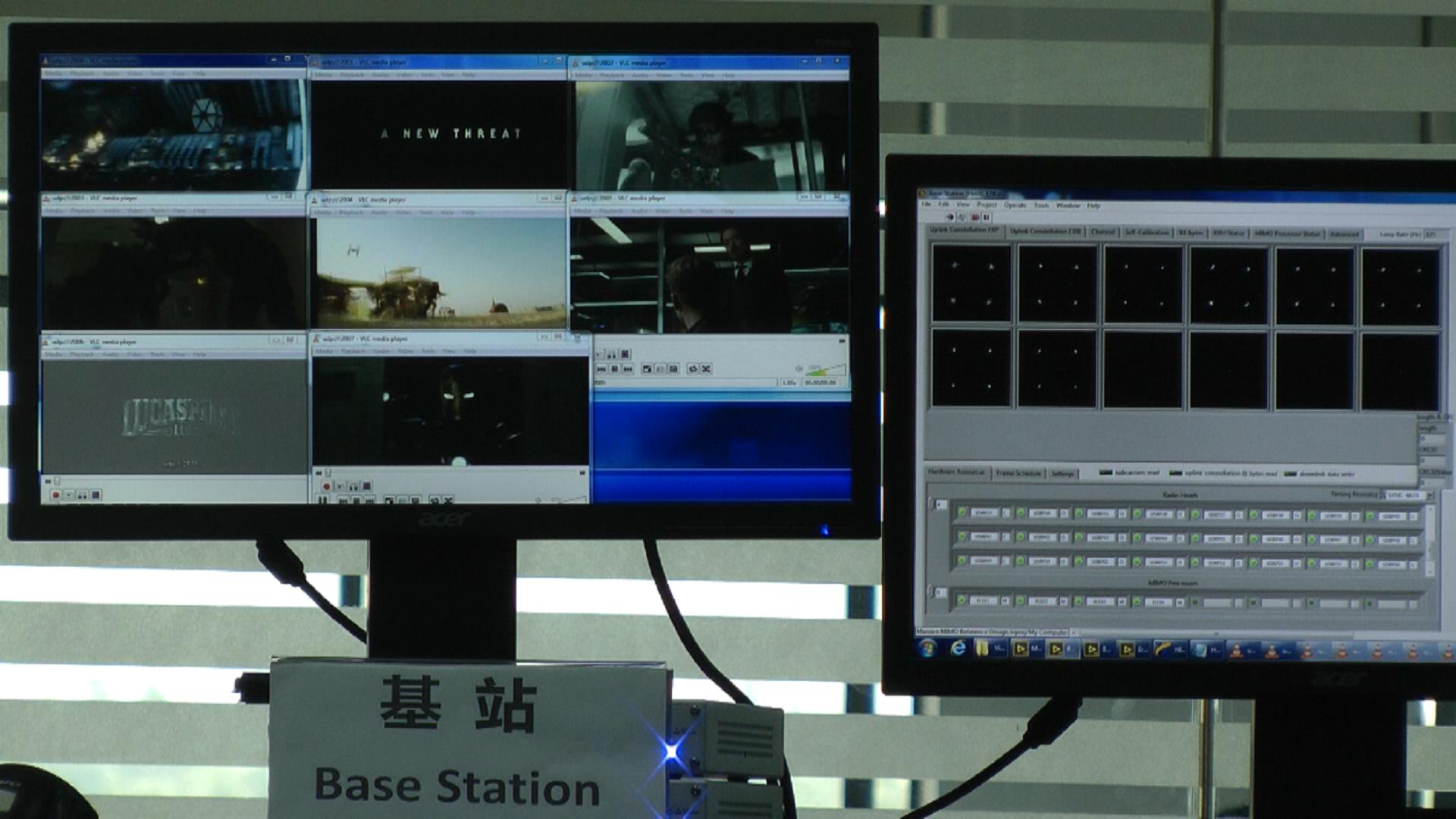}
\caption{Test results for uplink multiple video streaming transmission, the base station successfully recovered the multiple video streaming.}
\label{fig:17}
\end{figure}

\begin{figure}[htb]
\centering
\includegraphics[width=0.45\textwidth]{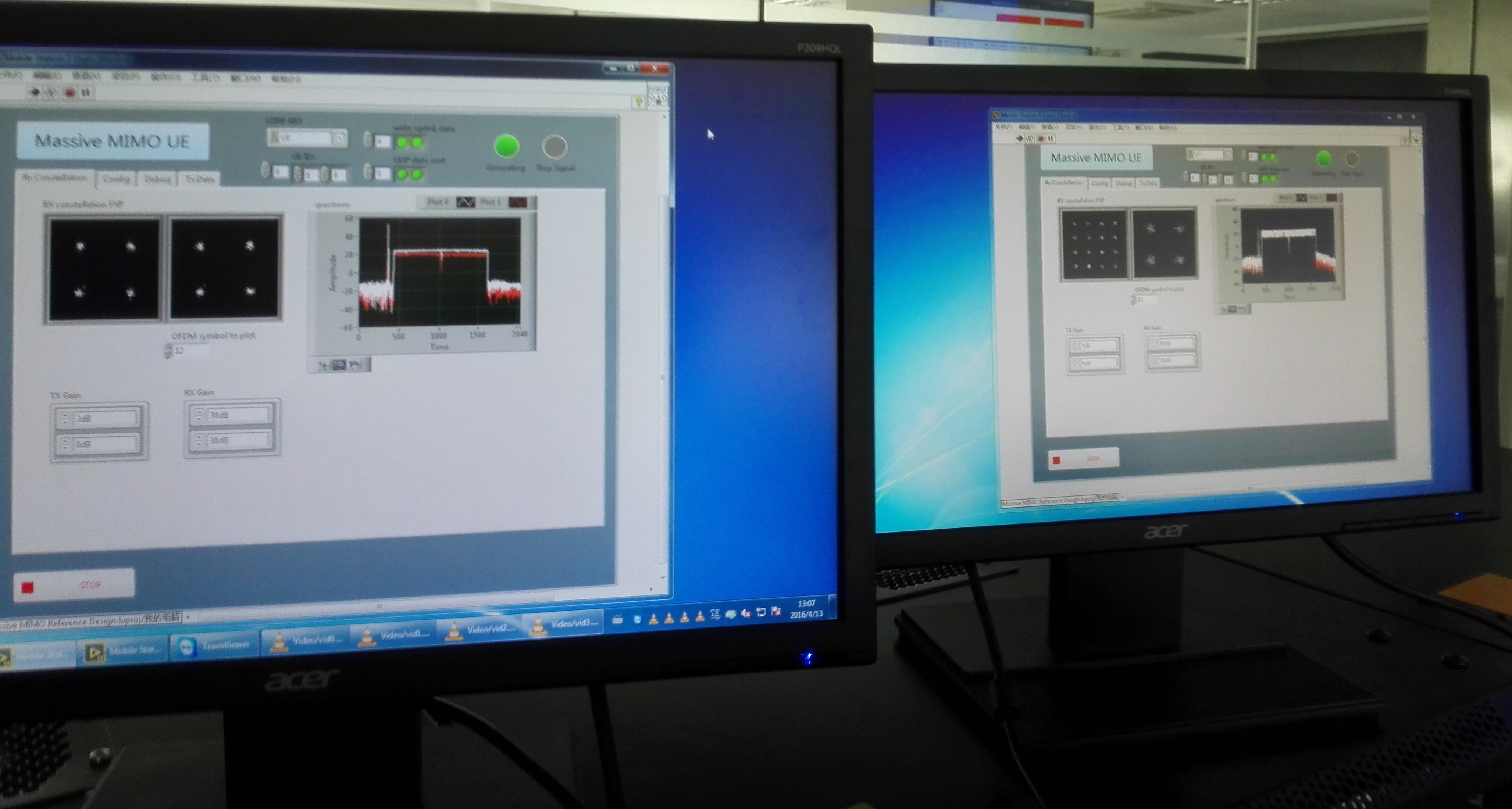}
\caption{Test results for downlink multi-user beamforming data transmission, the received signal spectrum and constellation are displayed by UEs.}
\label{fig:18}
\end{figure}

As a proof of concept, according to the designed frame structure, real-time uplink and downlink data transmission test is conducted.
In the transmission test, 8 single-antenna users transmit pilot and video stream data to the 128-antenna BS
with transmit power 15 dBm. %and system carrier frequency being 4.1 GHz.
Based on the received pilot signal, the BS performs channel estimation, MIMO detection and downlink precoding as described in system model in Section II, as well as uplink and downlink transmission procedure in Section IV.
The results of the test are showed in Fig. \ref{fig:17} and Fig. \ref{fig:18}.
From Fig. \ref{fig:17}, we can see QPSK indicated by the constellation in the right monitor is employed in the uplink,
and the base station successfully recovered the multiple video streams and displayed them in the left monitor, which validates the $268.8 Mbps$ peak rate achieved in the current bandwidth, modulation and user number configuration.
Higher modulation order can be configured flexibly if higher peak rate is needed.
Fig. \ref{fig:18} shows the received signal spectrum and recovered data by UEs in downlink.
Each constellation represents a single-antenna user and there are four single-antenna users presented in the figure
with three of them utilizing QPSK and one adopting 16-QAM (the other four users' results are the same and we do not present them for limited space).
The spectral efficiency we has achieved in current configuration is $16.8 bit/s/Hz$
and the maximum $80.64 bit/s/Hz$ can be obtained by the usage of 256-QAM and twelve single-antenna users.

\begin{figure}[htb]
\centering
\includegraphics[width=0.45\textwidth]{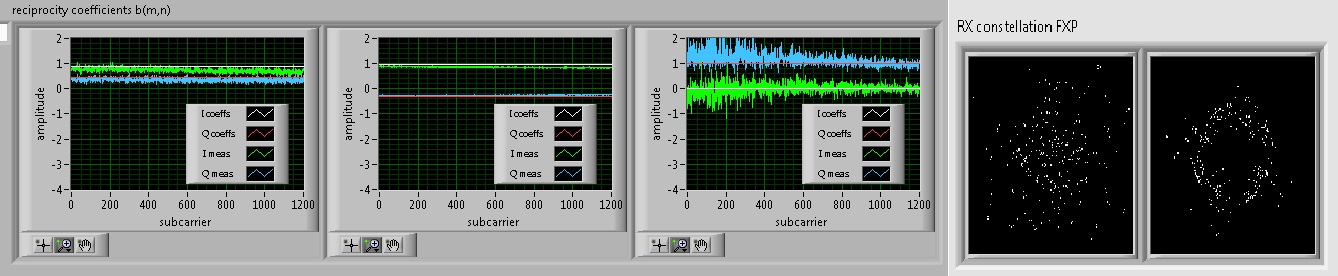}
\caption{Unsuccessful reciprocity calibration. Left: reciprocity calibration coefficients  when there is interference or the selected reference antenna is near the border of the antenna array, Right: the constellation of detected data for two single-antenna UEs in downlink.}
\label{fig:19}
\end{figure}

\begin{figure}[htb]
\centering
\includegraphics[width=0.45\textwidth]{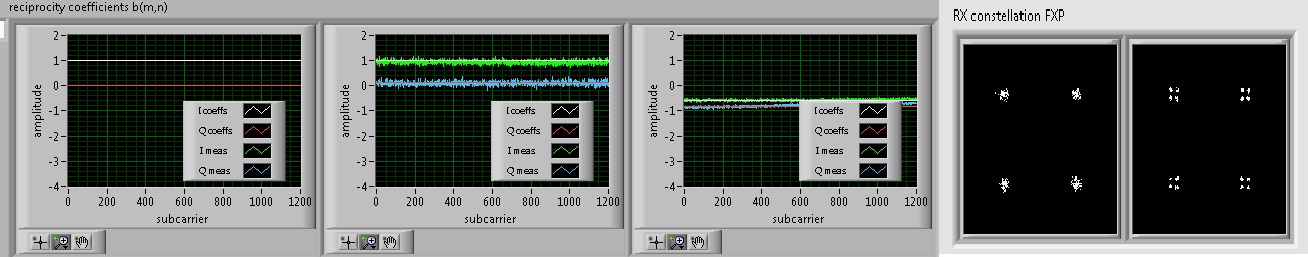}
\caption{Successful reciprocity calibration. Left: accurate reciprocity calibration coefficients, Right: the constellation of detected data for two single-antenna UEs in downlink.}
\label{fig:20}
\end{figure}

In order to verify the performance of the relative reciprocity calibration method,
we take several trials by setting different antennas as the reference antenna
or keep UEs transmitting during the calibration process.
The results shown in Fig. \ref{fig:19} and Fig. \ref{fig:20} imply that
when there is interference (e.g. UEs are transmitting signals) during the calibration process
or the selected reference antenna is near the border of the antenna array which leads to low SNR
for the antennas in the opposite side due to the array size,
the reciprocity calibration coefficients will be inaccurate and UEs can not recover the data they received in downlink because of the large interference between each other.
Note that the accurate calibration coefficients on the 1200 subcarriers (over a 20 MHz bandwidth)
depicted in Fig. \ref{fig:20} almost keep constant from the practical measurement,
and it may be beneficial to design the waveform used for reciprocity calibration.
Moreover, the geometry of antenna array also needs to be considered deliberately in reciprocity calibration methods.

%%%%%%%%%%%%%%%%%%%%%%%%%%%%%%%%%%%%%%%%%%%%%%%%%%%%%%%%%%%%%%%%%%%%%%%%%%%%%%%%%%%%%%%%%%%%%
\section{Conclusion}
In this paper, we have presented the design and implementation of a TDD-based 128-antenna massive MIMO
prototyping system supporting up to twelve single-antenna users on the same time-frequency resource block.
System model was provided to facilitate understanding of the prototyping system design.
Link-level simulation and the details of system hardware design were presented.
Both uplink and downlink data transmission were realized using the designed system architecture,
and several experiments, including multi-user massive MIMO channel measurement, multiple video streaming transmission in uplink, multi-user beamforming data transmission in downlink and so on, were carried out to test the system performance.
The results showed that 268.8 Mbps peak rate was achieved for eight single-antenna users under QPSK.
%and the maximum spectral efficiency would be 80.64 bit/s/Hz by the usage of 256-QAM and twelve single-antenna users.
In addition, reciprocity calibration has played an important role in multi-user downlink transmission
and needs to be designed with great care.
%which has already been our next plan in the near future along with DFT-based precoding.

% or


%\appendices

%\section{}

\section*{Acknowledgment}

The authors would like to thank Southeast University students Zijian Han and Feng Ji, Nanjing University of Posts and Telecommunications student Yu Yu, and Zhiya Information Technology engineers Xiaolong Miao and Wankai Tang for their assistance with the system architecture design and implementation.

\ifCLASSOPTIONcaptionsoff
  \newpage
\fi

\end{document}